\documentclass[showpacs,prc,preprint,nofootinbib,showkeys]{revtex4}
\pdfoutput=1

\usepackage{bm}
\usepackage{amsmath}
\usepackage{graphicx}
\usepackage{subfigure}
 
\usepackage[colorlinks=true,linktocpage=true,linkcolor=blue,citecolor=blue]{hyperref}
\usepackage[usenames]{color}

\newcommand{\bel}[1]{\begin{eqnarray}\label{#1}}
\newcommand{\eel}{\end{eqnarray}}
\newcommand{\rf}[1]{Eq.~(\ref{#1})}
\newcommand{\rfn}[1]{(\ref{#1})}

\def\l{\limits}

\newcommand{\s}{{\rm s}}		
\newcommand{\Qp}{{Q^+}}		
\newcommand{\Qm}{{Q^-}}		
\newcommand{\G}{{G}}		
\newcommand{\Qpm}{{Q^\pm}}		
\newcommand{\Q}{{Q}}		
\newcommand{\eq}{{\rm eq}}  
\newcommand{\an}{{\rm a}}  
\newcommand{\LRF}{{\rm LRF}}  
\newcommand{\teq}{\tau_\eq}

\newcommand{\bpT}{{\boldsymbol p}_T} 
 
\newcommand{\I}{{\cal I}}
\newcommand{\hI}{\hat{{\cal I}}}
\newcommand{\pT}{p_{T}}
\newcommand{\pL}{p_{L}}

\newcommand{\lp}{\left(}
\newcommand{\rp}{\right)}

\newcommand{\lsb}{\left[}
\newcommand{\rsb}{\right]}

\newcommand{\bea}{\begin{eqnarray}}
\newcommand{\eea}{\end{eqnarray}}
\newcommand{\beal}[1]{\begin{eqnarray}\label{#1}}
\newcommand{\eeal}{\end{eqnarray}}
\newcommand{\nn}{\nonumber}
\newcommand{\f}[2]{\frac{#1}{#2}}
\newcommand{\EQ}[1]{Eq.~(\ref{#1})}
\newcommand{\EQS}[1]{Eqs.~(\ref{#1})}
\newcommand{\EQSTWO}[2]{Eqs.~(\ref{#1})~and~(\ref{#2})}
\newcommand{\EQSM}[2]{Eqs.~(\ref{#1})--(\ref{#2})}
\newcommand{\EQB}[1]{(\ref{#1})}
\newcommand{\SEC}[1]{Sec.~\ref{#1}}

\newcommand{\APP}[1]{App.~\ref{#1}}

\newcommand{\p}{\partial}
\newcommand{\twpt}{(\tau,w,\pT)}
\newcommand{\tiwpt}{(\tau_0,w,\pT)}

\newcommand{\VP}{\vphantom{\frac{}{}}\!}

\begin{document}
 

\title{Coupled kinetic equations for quarks and gluons in the relaxation time approximation}

\author{Wojciech Florkowski} 
\affiliation{Institute of Nuclear Physics, Polish Academy of Sciences, PL-31342 Krak\'ow, Poland}
\affiliation{Institute of Physics, Jan Kochanowski University, PL-25406~Kielce, Poland} 

\author{Ewa Maksymiuk} 
\affiliation{Institute of Physics, Jan Kochanowski University, PL-25406~Kielce, Poland} 

\author{Radoslaw Ryblewski} 
\affiliation{Institute of Nuclear Physics, Polish Academy of Sciences, PL-31342 Krak\'ow, Poland}

\date{\today}


\begin{abstract}
Kinetic equations for quarks and gluons are solved numerically in the relaxation time approximation for the case of one-dimensional boost-invariant geometry. Quarks are massive and described by the Fermi-Dirac statistics, while gluons are massless and obey Bose-Einstein statistics. The conservation laws for the baryon number, energy, and momentum lead to two Landau matching conditions which specify the coupling between the quark and gluon sectors and determine the proper-time dependence of the effective temperature and baryon chemical potential of the system. The numerical results illustrate how a non-equlibrium mixture of quarks and gluons approaches hydrodynamic regime described by the Navier-Stokes equations with appropriate forms of the kinetic coefficients. The shear viscosity of a mixture is the sum of the shear viscosities of quark and gluon components, while the bulk viscosity is given by the formula known for a gas of quarks, however, with the thermodynamic variables characterising the mixture. Thus, we find that massless gluons contribute in a non-trivial way to the bulk viscosity of a mixture, provided quarks are massive. We further observe the hydrodynamization effect which takes place earlier in the shear sector than in the bulk one. The numerical studies of the ratio of the longitudinal and transverse pressures show, to a good approximation, that it depends on the ratio of the relaxation and proper times only. This behaviour is connected with the existence of an attractor solution for conformal systems. 

\end{abstract}



\maketitle 
%
\section{Introduction}
\label{sec:i}
%
Comparisons between predictions of hydrodynamic models and exact kinetic-theory results have become
an important method to verify the validity of hydrodynamic frameworks~\cite{Florkowski:2013lza,Florkowski:2013lya,Bazow:2013ifa,Florkowski:2014sfa,Florkowski:2014sda,Denicol:2014tha,Denicol:2014xca,Nopoush:2014qba,Florkowski:2015lra,Molnar:2016gwq,Martinez:2017ibh,Damodaran:2017ior} which are now our basic tools to interpret the processes of heavy-ion collisions studied experimentally at RHIC and the LHC~\cite{Romatschke:2009im,Ollitrault:2012cm,Gale:2013da,Jeon:2015dfa,Jaiswal:2016hex}. Such comparisons allow us also for deeper analyses of mutual relations between effective hydrodynamic models and microscopic, underlying theories~\cite{Noronha:2015jia,Heller:2016rtz,Denicol:2016bjh,Bemfica:2017wps}, for a recent review see~\cite{Florkowski:2017olj}. In this work we continue earlier studies on this topic and generalise previous results by studying a mixture of massive quarks and massless gluons forming a highly non-equilibrium system. Similarly to earlier works we restrict ourselves to boost-invariant systems~\cite{Bjorken:1982qr}.

Previous studies of mixtures~\cite{Florkowski:2012as,Florkowski:2013uqa,Florkowski:2014txa,Florkowski:2015cba} were restricted to the massless case and done mostly in the context of anisotropic hydrodynamics~\cite{Florkowski:2010cf,Martinez:2010sc}. In this paper we restrict ourselves to the kinetic-theory study, leaving an anisotropic hydrodynamics context for a separate investigation. Nevertheless, we use here the results of the first-order Navier-Stokes hydrodynamics to demonstrate the process of hydrodynamization of the system \cite{Heller:2011ju}. We find that the hydrodynamization in the shear sector (equalisation of the longitudinal, ${\cal P}_L$, and transverse, ${\cal P}_T$, pressures) takes place earlier than the hydrodynamization in the bulk sector (equalisation of the average and equilibrium pressures).  

In order to study the system behavior close to equilibrium we determine the shear and bulk viscosities of a mixture and find that the shear viscosity $\eta$ is simply a sum of the quark and gluon shear viscosities, $\eta = \eta_\Q + \eta_\G$. On the other hand,  the bulk viscosity of a mixture is given by the formula known for a massive quark gas, $\zeta$. Nevertheles, we find that $\zeta$ depends on thermodynamic coefficients characterising the whole mixture rather than quarks alone, which means that massless gluons contribute in a non-trivial way to the bulk viscosity (provided the quarks are massive).

 Interestingly, our studies of the time evolution of the ratio of the longitudinal and transverse pressures indicate that, to a very good approximation, it depends on the ratio of the relaxation and proper times only. This behaviour is related to the presence of an attractor which was found and discussed earlier for conformal systems~\cite{Heller:2015dha,Romatschke:2016hle,Spalinski:2016fnj,Romatschke:2017vte,Spalinski:2017mel,Strickland:2017kux}, and, quite recently, also for non-conformal ones \cite{Romatschke:2017acs}. 
 
The paper is organised as follows: In Secs.~\ref{sec:ke} and \ref{sect:motke} we introduce the system of kinetic equations for the quark-gluon mixture and study their momentum moments. This leads to two Landau matching conditions related to the baryon number and energy-momentum conservation. In Sec.~\ref{sect:tb} we discuss an algebraic method useful for dealing with tensors describing our main observables. This method is used in Secs.~\ref{sect:ie}, \ref{sec:ae}, and \ref{sect:e} to calculate
various thermodynamic variables for systems exhibiting isotropic, anisotropic, and exact distribution functions, respectively. In Sec.~\ref{sect:LM}, which is central analytic part of this work, we discuss the conservation laws and present two integral equations used to determine the proper-time dependence of the effective temperature and baryon chemical potential.  In Sec.~\ref{sect:res} we present our results describing proper-time dependence of various quantities, hydrodynamization process, and scaling properties of the ${\cal P}_L/{\cal P}_T$ ratio. We summarise and conclude in Sec.~\ref{sect:sumcon}. Appendices~\ref{s:tf}, \ref{s:NS} and \ref{s:shearbulk} contain: details of the calculations of the generalised thermodynamic functions, discussion of the Navier-Stokes equations, and the explicit calculation of the shear and bulk viscosities for a quark-gluon mixture, respectively.

\bigskip
In this work we use $x^\mu = (t,x,y,z)$ and  $p^\mu = (p^0 = E_p, p_x, p_y, p_z=\pL)$ to denote the particle space-time position and four-momentum. The longitudinal $(z)$ direction corresponds to the beam axis. The transverse momentum is $\pT = \sqrt{p_x^2 + p_y^2}$ and particles are assumed to be always on the mass shell,  $E_p = \sqrt{m^2 + \pT^2 + \pL^2}$. The scalar product of two four-vectors is  $a^\mu b_\mu = a^\mu  g_{\mu\nu} b^\nu \equiv a \cdot b$ where $g^{\mu\nu} = {\rm diag} \left(1,-1,-1,-1\right)$ is the metric tensor.  For the partial derivative we use the notation $\p_\mu\equiv \p/\p x^\mu$. Throughout the paper we use natural units with $c=k_B=\hbar=1$.

%
\section{Kinetic equations}
\label{sec:ke}
%
Our analysis is based on three coupled relativistic Boltzmann transport equations for quark, antiquark, 
and gluon phase-space distribution functions~$ f_{\s}(x,p)$ ~\cite{Florkowski:2012as,Florkowski:2013uqa,Florkowski:2014txa,Florkowski:2015cba},
\beal{kineq}
\lp p\cdot\p \rp  f_{\s}(x,p) &=&  {\cal C}\lsb  f_{\s}(x,p)\rsb,\quad \s=\Qp, \Qm,G \,. 
\eeal
The collisional kernel ${\cal C}$ in \rfn{kineq} is treated in the relaxation time approximation 
(RTA)~\cite{Bhatnagar:1954zz,Anderson:1974a,Anderson:1974b,Czyz:1986mr} 
\beal{colker}
{\cal C}\lsb f_\s(x,p)\rsb &=&   \lp p\cdot U \rp \frac{f_{\s, \eq}(x,p)-f_\s(x,p)}{\teq} \,,
\eeal
where $\teq$ is the relaxation time and the four-vector $U(x)$ describes the hydrodynamic flow. In numerical calculations we assume that $\teq$ is constant, which explicitly breaks conformal symmetry of the system. The other source of breaking of the conformal symmetry is a finite quark mass. The form of $U^\mu(x)$  in \rfn{kineq} is defined by choosing the Landau hydrodynamic frame. We note, however,  that for one-dimensional boost-invariant systems the structure of $U^\mu(x)$ follows directly from the symmetry arguments, see Sec. \ref{sect:bi}.

In \EQ{colker} the functions $f_{\s, \eq}(x,p)$ are standard equilibrium distribution functions, which (unless specified otherwise) take the Fermi-Dirac and Bose-Einstein forms for (anti)quarks and gluons, respectively,
\bea
f_{\Qpm, \eq}(x,p) &=& h^{+}_\eq\lp\f{  p\cdot U  \mp \mu }{T }\rp ,
\label{Qeq} \\
f_{\G, \eq}(x,p) &=& h^{-}_\eq\lp\f{p\cdot U}{T }\rp .
\label{Geq}
\eea
Here $T(x)$ is the effective temperature, $\mu(x)$ is the effective \textit{chemical potential of quarks}, and
\beal{feq}
h^{\pm}_\eq(a) &=&   \lsb\VP\exp(a) \pm 1 \rsb^{-1}.
\eeal
The same value of $T(x)$ appearing in Eqs.~(\ref{Qeq}) and (\ref{Geq}), as well as the same value of $\mu(x)$ appearing in the quark and antiquark distributions in Eq.~(\ref{Qeq}) introduces interaction between quarks, antiquarks and gluons -- all particles evolve toward the same local equilibrium defined by $T(x)$ and $\mu(x)$.  Since the baryon number of quarks is 1/3, we can use the relation %
\bel{muB}
\mu= \f{\mu_B}{3} \, ,
\eel
with $\mu_B$ being the baryon chemical potential. 

All particles are assumed to be on the mass shell, \mbox{$p^2 =p \cdot p=m^2$}, so that the invariant momentum measure is
\bea
 \int dP (\ldots)  \equiv  2  \int  d^4 p \, \Theta(p\cdot t) \delta(p^2-m^2) (\ldots)
 =    \int  \frac{d^3p}{E_p} (\ldots) \,,
\eea
where $\Theta$ is the Heaviside step function and $t^\mu$ is an arbitrary time-like four-vector.
Hereafter, the gluons are treated as massless, while quarks have a finite constant mass $m$.

%
\section{Moments of the kinetic equations}
\label{sect:motke}
%
We introduce the $n$-th moment operator in the momentum space 
\beal{mom}
\hI^{\mu_1\cdots\mu_n} (\dots) \equiv  \int \!dP\, p^{\mu_1}p^{\mu_2}\cdots p^{\mu_n} (\dots) \,,  
\eeal
with the zeroth moment operator defined as
\beal{mom0}
\hI (\dots) \equiv  \int \!dP\,(\dots) \,.
\eeal
Acting with $\hI^{\mu_1\cdots\mu_n}$ on the distribution functions $f_\s(x,p)$ and multiplying them by the degeneracy factors $k_\s$, one obtains the $n$-th moments of the distribution functions
\beal{momf}
\I_\s ^{\mu_1\cdots\mu_n} &\equiv& k_\s\, \hI^{\mu_1\cdots\mu_n} f_\s(x,p) \,.
\eeal
Here $k_\s\equiv g_\s/(2\pi)^3$, with $g_{Q^\pm}=3\times2\times N_f$ and $g_{G}=8\times2$ being the internal degeneracy factors for (anti)quarks  and gluons, respectively. In our calculations we assume that we deal with two (\emph{up} and \emph{down}) quark flavors  with equal mass, which reflects the SU(2) isospin symmetry. 

With the above definitions, the first and second moments of the distribution functions read 
\bea 
N_\s^\mu(x)&\equiv&\I_\s ^{\mu}  =k_\s   \int \!dP\, p^{\mu} f_\s(x,p) \,,
\label{ncurrs} \\
T_\s^{\mu\nu}(x)&\equiv&\I_\s ^{\mu\nu}  =   k_\s \int \!dP\, p^{\mu}p^{\nu} f_\s(x,p) \,,
\label{emtensors} 
\eea
which are identified with the \emph{particle number current} and the \emph{energy-momentum tensor} of the species ``$\s$'', respectively. In addition, we define the \emph{baryon number current}
\beal{bcurr}
B^\mu(x)&\equiv& \sum_\s q_\s\, N_\s^\mu(x) = \frac{k_{\Q^\pm}}{3}\int \!dP\, p^{\mu} \lsb\VP f_\Qp(x,p)-f_\Qm(x,p)\rsb, 
\eeal
where $q_\s=\left\{1/3,-1/3,0\right\}$ is the baryon number for quarks, antiquarks, and gluons, respectively. The total particle number current and total energy-momentum tensor read
\bea 
N^\mu(x)&=&\sum_\s N_\s^\mu(x) \,,
\label{totncurr} \\
T^{\mu\nu}(x)&=& \sum_\s T_\s^{\mu\nu}(x)\,.
\label{totemtensor} 
\eea

\bigskip
We now consider the $n$-th moments of the kinetic equations (\ref{kineq}), which are obtained by acting with the operator $\hI^{\mu_1\cdots\mu_n}$ given by (\ref{mom}) on their left- and right-hand sides and multiplying them by the degeneracy factors $k_\s$. The zeroth and first moments have the form
\bea
k_\s \hI  p^\mu \p_\mu  f_\s(x,p) &=& k_\s \hI p^\mu U_\mu\frac{f_{\s, \eq}(x,p)-f_\s(x,p)}{\teq},
\label{KEzerothmom} \\
k_\s \hI^{\nu} p^\mu \p_\mu  f_\s(x,p) &=& k_\s \hI^{\nu} p^\mu U_\mu\frac{f_{\s,\eq}(x,p)-f_\s(x,p)}{\teq},\,\,\,\,\,\,
\label{KEfirstmom} 
\eea
which, using Eqs.~(\ref{momf})--(\ref{emtensors}), may be rewritten as
\bea
\p_\mu {{N}}^\mu_\s  &=& U_\mu   \frac{{  N} ^\mu_{\s,\eq} -{  N} ^\mu_{\s }}{\teq},
\label{KEzerothmom2} \\
\p_\mu  {{T}}^{\mu\nu}_\s  &=& U_\mu   \frac{{ T}^{\mu\nu}_{\s,\eq} -{ T}^{\mu\nu}_{\s }}{\teq}.
\label{KEfirstmom2}  
\eea
Taking difference between $\s=\Qp$ and $\s=\Qm$ components of Eqs.~(\ref{KEzerothmom2}) we obtain the baryon current evolution equation
\beal{KEzerothmom3}
\p_\mu {{B}}^\mu   &=& U_\mu   \frac{{B}^\mu_\eq - {B}^\mu }{\teq} .
\eeal
On the other hand, when taking the sum over ``s''  components of Eqs.~(\ref{KEfirstmom2}) one gets the total energy and momentum conservation equation
\beal{KEfirsthmom3}
\p_\mu  {{T}}^{\mu\nu}  &=& U_\mu   \frac{{T}^{\mu\nu}_\eq -{T}^{\mu\nu}_{ }}{\teq}.
\eeal
In order to have the baryon number conserved it is required that the left-hand side of Eq.~(\ref{KEzerothmom3}) vanishes, $\p_\mu {{  B}}^\mu =0$. The latter implies vanishing of the right-hand side of Eq.~(\ref{KEzerothmom3}), which leads to the {\it Landau matching condition for baryon current}
\beal{baryonLMC}
 U_\mu  {B}^\mu_\eq =  U_\mu { B} ^\mu  .
\eeal
Analogously, the energy and momentum conservation means that the left-hand side of Eq.~(\ref{KEfirsthmom3}) vanishes, $\p_\mu  {{ T}}^{\mu\nu} = 0$. This condition results in vanishing of the right-hand side of Eq.~(\ref{KEfirsthmom3}), which leads to the {\it Landau matching condition for energy and momentum} 
\beal{emLMC}
U_\mu  {T}^{\mu\nu}_\eq =U_\mu {T}^{\mu\nu}.
\eeal
%
%
\section{Tensor decomposition}
\label{sect:tb}
%
It is convenient to introduce the four-vector basis~\cite{Florkowski:2011jg,Tinti:2013vba}
\bel{A}
\left( A_{(0)}, A_{(1)}, A_{(2)}, A_{(3)} \right) = \left( U,X,Y,Z \right),
\eel
which in the local rest frame (LRF) reads
\beal{eq:rfbasis}
&&A^\mu_{(0),{\rm LRF}} \equiv U^\mu_\LRF = (1,0,0,0), \nonumber \\
&&A^\mu_{(1),\LRF} \equiv X^\mu_\LRF = (0,1,0,0), \nonumber \\
&&A^\mu_{(2),\LRF} \equiv Y^\mu_\LRF = (0,0,1,0), \nonumber \\
&&A^\mu_{(3),\LRF} \equiv Z^\mu_\LRF = (0,0,0,1) \, .
\eeal
Using \EQS{eq:rfbasis} one may express the metric tensor as follows \cite{Martinez:2012tu}
\begin{equation}
g^{\mu \nu}=U^\mu  U^\nu  - \sum_{A\neq U} A^\mu  A^\nu  \, .
\label{eq:gbasis}
\end{equation}
The projector on the space orthogonal to the four-velocity, $\Delta^{\mu\nu}\equiv g^{\mu\nu} -U^\mu U^\nu$, takes the form
\begin{equation}
\Delta^{\mu \nu} =  - \sum_{A\neq U} A^\mu  A^\nu \, ,
\label{eq:transproj}
\end{equation}
and satisfies the conditions $U_{\mu} \Delta^{\mu\nu} = 0$, $\Delta^{\mu}_{\,\,\,\alpha}\Delta^{\alpha\nu}=\Delta^{\mu\nu}$ and $\Delta^{\mu}_{\,\,\,\mu}=3$.   
The basis (\ref{eq:rfbasis}) is a unit one in the sense that 
\begin{eqnarray}
A \cdot B =
\begin{cases}
0   & \hbox{for   } A\neq B, \\
1   & \hbox{for   } A= B=U,   \\
-1  & \hbox{for   } A= B\neq U,   
\end{cases}
  \label{eq:orthonormal} 
\end{eqnarray}
and complete so that any four-vector may be decomposed in the basis $A_{(\alpha)}$.~In particular, one may express the particle number flux as follows
\beal{eq:nfdecomp}
N_\s^\mu(x) &=& \sum_A  n_{A}^{\s} A_{ }^\mu, 
\eeal
where the coefficients $n_{A}^{\s}$, due to Eqs.~(\ref{eq:orthonormal}), are given by the projections
\beal{eq:nfdecompcoef}
n_{A}^{\s} &=&   A_{\mu}^{\,}  N_\s^\mu(x)  \, A^2, 
\eeal
with $A^2 = A \cdot A$ (note that $A^2=-1$ for space-like four-vectors of the basis (\ref{eq:rfbasis})). The tensorial basis for the rank-two tensors is constructed using tensor products of the basis four-vectors $A^\mu_{(\alpha)}$. Thus the decomposition of the energy-momentum tensor takes the form
\beal{eq:emtdecomp}
T_\s^{\mu\nu}(x) &=& \sum_{A,B} t_{AB}^{\s} A^\mu B^\nu,
\eeal
with the components of $T_\s^{\mu\nu}(x)$ defined in the following way
\beal{eq:emtdecompcoef}
 t_{AB}^{\s} &=& A_{\mu}^{\,} B_{\nu}^{\,} T_\s^{\mu\nu}(x) \, A^2  B^2.
\eeal
Using Eqs.~(\ref{ncurrs}) and (\ref{emtensors})  in Eqs.~(\ref{eq:nfdecompcoef}) and (\ref{eq:emtdecompcoef}) in one gets
\bea 
n_{A}^{\s} &=&  k_\s   \int \!dP\, \lp p \cdot A  \rp A^2 f_\s(x,p)   ,
\label{eq:nfdecompcoefgen}\\
t_{AB}^{\s} &=& k_\s\int \!dP\,  \lp p \cdot A \rp \lp p \cdot B \rp A^2 B^2 f_\s(x,p)  .
\label{eq:emtdecompcoefgen}
\eeal
%
%
\section{Isotropic distributions}
\label{sect:ie}
%
In the case of momentum-isotropic distribution functions (in particular, in the case of equilibrium distribution functions $f_{\s, \eq}(x,p) =  f_{\s, \eq}(p\cdot U(x))$, as defined by Eqs.~(\ref{Qeq}) and (\ref{Geq})), which are invariant with respect to $SO(3)$ rotations  in the three-momentum space, by the symmetry  of the integrands  in Eqs.~(\ref{eq:nfdecompcoefgen}) and (\ref{eq:emtdecompcoefgen}) one has
\begin{eqnarray}
 n_{A}^{\s, \eq} & = &  k_\s   \int \!dP\, \lp p\cdot A \rp A^2 f_{\s, \eq}   =0 \quad \hbox{if     } A\neq U,
\label{isonfcoefgen}\\
t_{AB}^{\s, \eq} & = & k_\s\int \!dP\, \lp p\cdot A \rp \lp p\cdot B \rp A^2 B^2 f_{\s, \eq} =0  \quad \hbox{if     } A\neq B,
\label{isoemtcoefgen}
\end{eqnarray}
so that for the momentum-isotropic state Eqs.~(\ref{eq:nfdecomp}) and (\ref{eq:emtdecomp}) have the following structure
\begin{eqnarray} 
N_{\s, \eq}^\mu(x)&=&  {\cal N}^{\s, \eq} U^\mu, \label{ncurreq} \\
T_{\s, \eq}^{\mu\nu}(x)&=& {\cal E}^{\s, \eq} U^\mu U^\nu -{\cal P}^{\s, \eq} \Delta^{\mu\nu},
\label{emtensoreq} 
\end{eqnarray} 
with 
\begin{equation}
{\cal N}^{\s, \eq} = n_{U}^{\s, \eq}, \quad {\cal E}^{\s, \eq} = t_{UU}^{\s, \eq}, 
\quad {\cal P}^{\s, \eq} = t_{XX}^{\s, \eq} = t_{YY}^{\s, \eq} = t_{ZZ}^{\s, \eq},
\label{eqvars}
\end{equation}
being the particle density, energy density, and pressure in equilibrium. Explicit forms of these expressions are given in \APP{ss:ie}.
%
\section{Anisotropic Romatschke-Strickland distributions}
\label{sec:ae}
%
It is also useful to consider anisotropic phase-space distributions introduced by Romatschke and Strickland in~\cite{Romatschke:2003ms}. In the covariant form they read \cite{Florkowski:2012as}
\begin{eqnarray}
f_{\Qpm, \an}(x,p) \!\!\!&=&\!\!\! h^{+}_\eq\lp\f{\sqrt{\lp p \cdot U\rp^2+\xi_\Q  \lp p \cdot Z\rp^2} \mp \lambda }{\Lambda_\Q}\rp,
\,\,\,\,  \label{Qa} \\
f_{\G, \an}(x,p) \!\!\!&=&\!\!\! h^{-}_\eq\lp\f{\sqrt{\lp p \cdot U\rp^2+\xi_\G  \lp p \cdot Z\rp^2}}{\Lambda_G}\rp,
\label{Ga}
\end{eqnarray}
where $\xi_\Q(x)=\xi_\Qp(x)=\xi_\Qm(x)$ is the quark anisotropy parameter, $\Lambda_\Q(x)=\Lambda_\Qp(x)=\Lambda_\Qm(x)$ is the quark transverse-momentum scale, and $\lambda(x)$ is the non-equilibrium baryon chemical potential of quarks. Similarly,
$\xi_G(x)$ is the gluon anisotropy parameter and $\Lambda_G(x)$ is the gluon transverse-momentum scale. The anisotropy parameters $\xi_{\rm s}$ vary in the range $-1 < \xi_{\rm s} <  \infty$, with the cases $-1 < \xi_{\rm s} < 0$, $0 < \xi_{\rm s} < \infty$ and $\xi_{\rm s}=0$ corresponding to the prolate, oblate and isotropic momentum distribution, respectively.

The distributions defined by Eqs.~(\ref{Qa}) and (\ref{Ga}) are invariant only with respect to $SO(2)$ rotations around the $z$ direction in the three-momentum space. In this case one still has 
\begin{eqnarray}
 n_{A}^{\s, \an} &=&  0 \quad \hbox{if     } A\neq U,
\label{anisonfcoefgen}\\
t_{AB}^{\s, \an} &=& 0  \quad \hbox{if     } A\neq B,
\label{anisoemtcoefgen}
\end{eqnarray}
and Eqs.~(\ref{ncurrs}) and (\ref{emtensors}) have the following structure~\cite{Florkowski:2008ag}
\begin{eqnarray} 
N_{\s,\an}^\mu(x)&=&  {\cal N}^{\s,\an} U^\mu, \label{ncurran} \\
T_{\s,\an}^{\mu\nu}(x)&=& {\cal E}^{\s,\an} U^\mu U^\nu -{\cal P}^{\s,\an}_T \Delta_T^{\mu\nu} +{\cal P}^{\s,\an}_L Z^\mu Z^\nu,
\label{emtensoran} 
\end{eqnarray} 
with 
\beal{anisovars}
{\cal N}^{\s,\an} &=& n_{U}^{\s, \an}, \quad {\cal E}^{\s,\an} =  t_{UU}^{\s, \an},  
\quad {\cal P}^{\s,\an}_T = t_{XX}^{\s, \an}= t_{YY}^{\s, \an}, \quad {\cal P}^{\s,\an}_L = t_{ZZ}^{\s, \an} .
\eeal
Here $\Delta_T^{\mu\nu} = -\left(X^\mu X^\nu + Y^\mu Y^\nu \right)$ is the projection operator orthogonal to $U$ and $Z$. Explicit forms of \EQS{anisovars} are given in \APP{ss:ae}.

\section{Exact solutions of the kinetic equations}
\label{sect:e}
%
In order to solve Eqs.~(\ref{baryonLMC}) and (\ref{emLMC}) we need to know the form of the distribution functions $f_{\s}(x,p)$ being solutions of the kinetic equations~(\ref{kineq}). In general, such solutions are difficult to find and Eqs.~(\ref{kineq}) may be at best solved numerically. However, it is possible to find \textit{formal} analytic solutions of Eqs.~(\ref{kineq}) in the case where the system is boost invariant and transversally homogeneous. Below, we discuss this case in more detail.

\subsection{Boost-invariance and transversal homogeneity}
\label{sect:bi}
%
Hereafter, we assume that the considered system is boost-invariant in the longitudinal (beam) direction and homogeneous in the transverse direction. In such a case we may choose~\cite{Florkowski:2011jg}
\begin{eqnarray}
 U^\mu &=& (t/\tau,0,0,z/\tau),  \\
 X^\mu &=& (0,1,0,0),   \\
 Y^\mu &=& (0,0,1,0),   \\
 Z^\mu &=& (z/\tau,0,0,t/\tau),
 \label{UZbinv}
\end{eqnarray}
where $\tau$ is the (longitudinal) proper time
\begin{eqnarray}
\tau = \sqrt{t^2-z^2}.
\end{eqnarray}
As a result the system becomes effectively one-dimensional. Since its evolution is governed completely by the proper time that mixes $t$ and $z$, one usually refers to such a system as  (0+1)-dimensional.
%
\subsection{Boost-invariant Bialas-Czyz variables}
\label{sect:wv}
%
In the case of (0+1)-dimensional system exhibiting symmetries discussed in the previous section it is convenient to use the variables $w$ and $v$ which are defined as follows  \cite{Bialas:1984wv,Bialas:1987en}
\bea 
w & =& t \pL - z E_p =  - \, \tau \, p \cdot Z, \label{w}\\
v  &=& t  E_p - z \pL = \tau \, p \cdot U.  \label{v}
\eea 
Due to the fact that particles are on the mass shell $w$ and $v$ are related by the formula
\beal{v1}
v\twpt &=&   \sqrt{w^2+\lp m^2+\pT^{\,2}\rp  \tau^2}.  
\eeal
Equations (\ref{w}) and (\ref{v}) can be inverted to express the energy and longitudinal momentum of a particle in terms of $w$ and $v$, namely
\begin{equation}   
E_p= \f{vt+wz}{\tau^2},\quad \pL=\f{wt+vz}{\tau^2}.  
\label{p0p3}
\end{equation}
The  Lorentz invariant momentum-integration measure can be written now as
\begin{eqnarray}
dP = \f{d^3p}{E_p} = \f{dw\,d^2\pT}{v} .
\label{dP}
\end{eqnarray}
For boost-invariant systems, all scalar functions of space and time, such as the effective temperature $T$ and quark chemical potential $\mu$, may depend only on $\tau$. In addition, one can check that the phase-space distribution functions (which are Lorentz scalars) may depend only on the variables $w$, $\tau$ and $\bpT$. We use these properties in the next section.
%
\subsection{Formal solutions of the kinetic equations}
\label{sect:boost-inv-eq}
%
With the help of the variables $w$, $v$ and $\bpT$ we can rewrite~(\ref{kineq})  in a simple form~\cite{Florkowski:2013lza,Florkowski:2013lya,Baym:1984np,Baym:1985tna}
\beal{BIke}
\frac{\p f_\s \left(\tau, w,  \bpT\right)}{\p\tau}   &=& \f{f_{\s,\eq}\left(\tau, w, \pT\right)-f_\s \left(\tau, w,  \bpT\right)}{\teq},  
\eeal
where the boost-invariant versions of the equilibrium distribution functions are straightforward to find using  (\ref{w}) and (\ref{v})  
\begin{eqnarray}
f_{\Qpm, \eq} \twpt &=&h^{+}_\eq \lp\f{\sqrt{\lp\f{w}{\tau}\rp^2+ \pT^2 + m^2  }\mp \mu}{T } \,  \rp  , 
\label{BIQeq0} \\
 f_{\G, \eq} \twpt &=&
 h^{-}_\eq\lp\f{\sqrt{\lp\f{w}{\tau}\rp^2+\pT^2}}{T } \,  \rp . 
\label{BIGeq0}
\end{eqnarray}
Below we assume that distribution functions $f_\s\left(\tau, w,  \bpT\right)$ are even functions of $w$, and depend only on the magnitude of $\bpT$ \footnote{In our analysis we restrict ourselves to the initial distributions in the RS form, which are  SO(2) invariant in transverse momentum space and thus depend only on the magnitude of $\bpT$. },
\begin{eqnarray}
f_\s\twpt = f_\s(\tau,-w,\pT)  .
\label{symoff}
\end{eqnarray}
The formal solutions of \EQS{BIke} have the form~\cite{Florkowski:2013lza,Florkowski:2013lya,Baym:1984np,Baym:1985tna}
\begin{eqnarray}
f_\s\twpt &=& D(\tau,\tau_0) f_\s^0(w,\pT) + \int\l_{\tau_0}^{\tau} \f{d \tau'}{\teq^\prime}\  D(\tau,\tau') 
f_{\s,\eq}(\tau^\prime,w,p_T). \label{formsolQ}  
\end{eqnarray}
where $f_\s^0(w,p_T)\equiv f_\s (\tau_0,w,p_T)$ is the initial distribution function (we have introduced here the notation
$\teq^\prime = \teq(\tau^\prime)$ for the general case where the equilibration time may depend on the proper time).

\subsection{Damping function}
%
In \EQ{formsolQ} we have introduced the damping function
\begin{equation}
D(\tau_2,\tau_1)=\mathrm{exp}\Bigg[ -\int\l_{\tau_1}^{\tau_2} \f{d\tau''}{\teq(\tau'')}\Bigg].
\label{Damp1}
\end{equation}
The function $D(\tau_2,\tau_1)$ satisfies the two differential relations
\begin{eqnarray}
\f{\p D(\tau_2,\tau_1)}{\p \tau_2}  = - \f{D(\tau_2,\tau_1)}{\teq(\tau_2)}, \quad
\f{\p D(\tau_2,\tau_1)}{\p \tau_1}  =  \f{D(\tau_2,\tau_1)}{\teq(\tau_1)}
\label{Damp2}
\end{eqnarray}
and converges to unity if the two arguments are the same, $D(\tau,\tau)=1$. These properties imply the identity~\cite{Florkowski:2014txa}
\begin{eqnarray}
1 = D(\tau,\tau_0) + \int\l_{\tau_0}^\tau \f{d\tau^\prime}{\teq(\tau^\prime)}
D(\tau,\tau^\prime).
\label{Damp3}
\end{eqnarray}
For a constant relaxation time used in this work Eq.~(\ref{Damp1}) reduces to
\begin{equation}
D(\tau_2,\tau_1)=\exp\lsb-\f{\tau_2-\tau_1}{\teq}\rsb .
\label{Damp2a}
\end{equation} 

\subsection{Initial distributions}

In what follows we assume that the initial distributions $f_\s^0(w,p_T)$ are given by the anisotropic Romatschke-Strickland (RS) forms $f_{\s,\an}^0(w,p_T)$ which follow from Eqs.~(\ref{Qa}) and (\ref{Ga}),
\begin{eqnarray}
&& f_{\Qpm, \an} \tiwpt = 
h^{+}_\eq \lp\f{\sqrt{\lp1+\xi_\Q^0\rp\lp\frac{w}{\tau_0}\rp^2+  m^2+\pT^2   } \mp \lambda ^0 }{\Lambda_\Q^0}   \rp  ,
\label{BIQeq} \\
&& f_{\G, \an} \tiwpt = h^{-}_\eq\lp\f{\sqrt{\lp1+\xi_\G^0\rp\lp\frac{w}{\tau_0}\rp^2+\pT^2   }}{\Lambda_\G^0} \,  \rp . \nn 
\label{BIGeq}
\end{eqnarray}
Here $\xi_\s^0\equiv\xi_\s(\tau_0)$, $\Lambda_\s^0\equiv\Lambda_\s(\tau_0)$,  and $\lambda^0\equiv\lambda(\tau_0)$ are initial parameters.

In view of the form~(\ref{formsolQ}), the use of Eqs.~(\ref{Qa}) and (\ref{Ga}) implies that the decomposition of the particle current and the energy-momentum tensor for \rfn{formsolQ} has the form of \EQSTWO{ncurran}{emtensoran}, namely
\begin{eqnarray} 
N_{\s}^\mu(x)&=&  {\cal N}^{\s} U^\mu, \label{ncurr} \\
T_{\s}^{\mu\nu}(x)&=& {\cal E}^{\s} U^\mu U^\nu -{\cal P}^{\s}_T \Delta_T^{\mu\nu} +{\cal P}^{\s}_L Z^\mu Z^\nu,
\label{emtensor} 
\end{eqnarray} 
with 
\beal{svars}
{\cal N}^{\s} &=& n_{U}^{\s}, \quad {\cal E}^{\s} = t_{UU}^{\s},  
\quad {\cal P}^{\s}_T = t_{XX}^{\s}= t_{YY}^{\s},  \quad {\cal P}^{\s}_L = t_{ZZ}^{\s}.  
\eeal
Hereafter, we refer to results obtained with the solution $(\ref{formsolQ})$ as the \textit{kinetic} or \textit{exact} ones.

\section{Baryon number and four-momentum conservation}
\label{sect:LM}

\subsection{Baryon number conservation}
%
Using the expression for the baryon number current (\ref{bcurr}) and the decompositions (\ref{ncurreq}) and (\ref{ncurr}) one may rewrite \EQ{baryonLMC} as 
\begin{equation}
{\cal B}^\eq = {\cal B},
\label{BIbaryonLMC}
\end{equation} 
where we define the equilibrium and exact baryon number densities as
\begin{equation}
{\cal B}^\eq = \f{1}{3} \left( {\cal N}^{\Qp, \eq} -{\cal N}^{\Qm, \eq} \right), 
\qquad
{\cal B}  = \f{1}{3} \left( {\cal N}^{\Qp } -{\cal N}^\Qm \right).
\end{equation}
The explicit formula for ${\cal B}^\eq(\tau)$ is derived in \APP{ss:ie}, see Eq.~(\ref{BeqApp}),
\begin{equation}
{\cal B}^\eq(\tau) = \f{16 \pi k_\Q T^3}{3} \sinh\lp\frac{\mu}{T}\rp\,{\cal H}_{\cal B}\lp \frac{m}{T},   \frac{\mu}{T}\rp,
\label{FIRST-EQUATION-0}
\end{equation}
where the function ${\cal H}_{\cal B}$ is defined by Eq.~(\ref{HB}). The formula for ${\cal B}(\tau)$ is more complicated and is given in \APP{ss:esotke}, see Eq.~(\ref{BApp}).
It contains an integral over the time history of the functions $T' \equiv T(\tau')$ and $\mu'\equiv\mu(\tau')$
in the range $\tau_0 \leq \tau' \leq \tau$. Consequently, Eq.~(\ref{BIbaryonLMC}) becomes
an integral equation 
\begin{widetext}
\begin{eqnarray}
T^3 \sinh\lp\frac{\mu}{T}\rp\,{\cal H}_{\cal B}\lp \frac{m}{T},   \frac{\mu}{T}\rp  
&=&   \f{\tau_0 \lp\Lambda_\Q^0\rp^3}{\tau \sqrt{1+\xi_\Q^0}} \sinh\lp\frac{\lambda^0}{\Lambda_\Q^0}\rp\,{\cal H}_{\cal B}\lp \frac{m}{\Lambda_\Q^0},\frac{\lambda^0}{\Lambda_\Q^0} \rp D(\tau,\tau_0) \nn \\
&& + \int\l_{\tau_0}^{\tau} \f{d \tau'}{\teq^\prime}\  D(\tau,\tau')  \f{\tau^\prime \lp T^\prime\rp^3}{\tau  } \sinh\lp\frac{\mu^\prime}{T^\prime}\rp\,{\cal H}_{\cal B}\lp \frac{m}{T^\prime}, \frac{\mu^\prime}{T^\prime}\rp .
\label{FIRST-EQUATION}
\end{eqnarray} 
\end{widetext}
 Equation (\ref{FIRST-EQUATION}) is a single equation for two functions, $T(\tau)$ and $\mu(\tau)$. The second necessary equation required 
 for their determination is obtained from the Landau matching condition for the energy, which we discuss in the next section.

Meanwhile, it is interesting to notice that Eq.~(\ref{FIRST-EQUATION}) can be rewritten as an integral equation for the function ${\cal B}(\tau)$, namely
\begin{eqnarray}
{\cal B}(\tau) = \f{\tau_0}{\tau} {\cal B}(\tau_0) D(\tau,\tau_0)
+ \int\l_{\tau_0}^{\tau} \f{d \tau'}{\teq^\prime} \f{\tau'}{\tau}  {\cal B}(\tau') \, D(\tau,\tau').
\label{FIRST-EQUATION-1}
\end{eqnarray}
By differentiating (\ref{FIRST-EQUATION-1}) with respect to $\tau$ we get
\begin{equation}
\f{d{\cal B}(\tau)}{d\tau} +\f{{\cal B}(\tau)}{\tau} = 0,
\label{FIRST-EQUATION-2}
\end{equation}
which is nothing else but the form of baryon number conservation law valid for the Bjorken geometry (in the original Bjorken paper \cite{Bjorken:1982qr}
the same equation was obtained for the conserved entropy current). Equation~(\ref{FIRST-EQUATION-2}) has scaling solution 
\begin{equation}
{\cal B}(\tau) = \f{\tau_0}{\tau} \, {\cal B}(\tau_0).
\label{FIRST-EQUATION-3}
\end{equation}
Combining   (\ref{BIbaryonLMC}) and (\ref{FIRST-EQUATION-0}) with (\ref{FIRST-EQUATION-3}) we find the equation
\begin{equation}
\sinh\lp\frac{\mu}{T}\rp\,{\cal H}_{\cal B}\lp \frac{m}{T},   \frac{\mu}{T}\rp 
= \f{3 \tau_0 {\cal B}(\tau_0)}{16 \pi k_\Q \tau T^3},
\label{FIRST-EQUATION-4}
\end{equation}
which allows to determine $\mu$ in terms of $T$ and $\tau$ for a given initial baryon number density. Unfortunately, in the general case we study (Fermi-Dirac statistics for quarks) Eq.~(\ref{FIRST-EQUATION-4}) is an implicit equation for $\mu$. The situation simplifies in the case of classical statistics, where the function ${\cal H}_{\cal B}$ becomes independent of $\mu$.

\subsection{Four-momentum conservation}

Using the expression for the energy-momentum tensor (\ref{totemtensor}) and the decompositions (\ref{emtensoreq}) and (\ref{emtensor}) one may rewrite \EQ{emLMC} as 
\begin{equation}
{\cal E}^\eq = {\cal E},
\label{energyLMC}
\end{equation} 
where ${\cal E}^\eq$ and ${\cal E}$ contain contributions from quarks, antiquarks, and gluons
\begin{equation}
{\cal E}^\eq =  {\cal E}^{\Qp, \eq} + {\cal E}^{\Qm, \eq}  + {\cal E}^{G, \eq} ,
\end{equation}
\begin{equation}
{\cal E}  = {\cal E}^{\Qp } +{\cal E}^\Qm  + {\cal E}^G.
\end{equation}
Using Eqs.~(\ref{eq:eqeq}), (\ref{eq:egeq}), (\ref{eq:eq}), and (\ref{eq:eg}) we obtain
\bigskip
\begin{eqnarray}
&&  T^4 \lsb \tilde{{\cal H}}^+\lp 1, \frac{m}{T}, - \frac{\mu}{T}\rp +
 \tilde{{\cal H}}^+\lp 1, \frac{m}{T}, + \frac{\mu}{T}\rp  + r  \tilde{{\cal H}}^-\lp 1, 0,0\rp \rsb \nn \\
&&  =     \lp\Lambda_\Q^0\rp^4 
\lsb \tilde{{\cal H}}^+\lp \f{\tau_0}{\tau \sqrt{1+\xi_\Q^0}}, \frac{m}{\Lambda_\Q^0}, - \frac{\lambda^0}{\Lambda_\Q^0}\rp 
+ \tilde{{\cal H}}^+\lp \f{\tau_0}{\tau \sqrt{1+\xi_\Q^0}}, \frac{m}{\Lambda_\Q^0}, + \frac{\lambda^0}{\Lambda_\Q^0}\rp \rsb
D(\tau,\tau_0) \nn \\
&& +    \int\l_{\tau_0}^{\tau} \f{d \tau'}{\teq^\prime}\  D(\tau,\tau') \lp T^\prime\rp^4
\lsb \tilde{{\cal H}}_{\cal  }^+\lp \f{\tau^\prime}{\tau  }, \frac{m}{T^\prime}, - \frac{\mu^\prime}{T^\prime}\rp  +
\tilde{{\cal H}}_{\cal  }^+\lp \f{\tau^\prime}{\tau  }, \frac{m}{T^\prime}, + \frac{\mu^\prime}{T^\prime}\rp   \rsb
\label{SECOND-EQUATION} \\
&& +   r  \lsb  \lp\Lambda_\G^0\rp^4 \tilde{{\cal H}}^-\lp \f{\tau_0}{\tau \sqrt{1+\xi_\Q^0}},  0,0\rp D(\tau,\tau_0) 
+   \int\l_{\tau_0}^{\tau} \f{d \tau'}{\teq^\prime}\  D(\tau,\tau') \lp T^\prime\rp^4 \tilde{{\cal H}}_{\cal  }^-\lp \f{\tau^\prime}{\tau  }, 0,0\rp\rsb, \nn
\end{eqnarray}
where the functions $\tilde{{\cal H}}^\pm$ are defined by Eqs.~(\ref{Ht}) and $r$ is the ratio of the degeneracy factors
\begin{equation}
r = \f{k_\G }{k_\Q} = \f{g_\G }{g_\Q} = \f{4}{3}.
\end{equation}

Equations (\ref{FIRST-EQUATION}) and (\ref{SECOND-EQUATION}) are two integral equations that are sufficient to determine the proper-time dependence of the functions $T(\tau)$ and $\mu(\tau)$. This is done usually by the iterative method~\cite{Banerjee:1989by}. The two initial, to large extent arbitrary, input functions  $T_{\rm in}(\tau)$ and $\mu_{\rm in}(\tau)$ are used on the right-hand sides of (\ref{FIRST-EQUATION}) and (\ref{SECOND-EQUATION})   and the new values  $T_{\rm out}(\tau)$ and $\mu_{\rm out}(\tau)$ are calculated from the left-hand sides. They are next used as $T_{\rm in}(\tau)$ and $\mu_{\rm in}(\tau)$ on the right-hand sides to get updated values of  $T_{\rm out}(\tau)$ and $\mu_{\rm out}(\tau)$. Such procedure is repeated until the updated values agree well with the initial values. We have found that the stable results are obtained with about 50 iterations if the final proper time is 5.0 fm. The time of the calculations grows quadratically with the final proper time. 

Our use of the two coupled integral equations is similar to the case studied previously in~\cite{Florkowski:2016qig}. We find that it is more straightforward than using (\ref{SECOND-EQUATION}) together with (\ref{FIRST-EQUATION-4}). However, the situation is different in the case of classical statistics, where (\ref{FIRST-EQUATION-4}) can be used to determine analytically $\mu/T$. In this case, the expression for $\mu/T$ obtained from (\ref{FIRST-EQUATION-4}) may be substituted into (\ref{SECOND-EQUATION}) and we are left with a single integral equation for the function $T(\tau)$.

One may check, using \rfn{dH2} and \rfn{dtH2}, that \rf{SECOND-EQUATION} is consistent with the formula
\bel{SECOND-EQUATION-1}
\f{d{\cal E}  }{d\tau} = -\f{{\cal E}  + {\cal P}_L  }{\tau},
\eel
where ${\cal P}_L  = {\cal P}_L^{\Qp } +{\cal P}_L^\Qm  + {\cal P}_L^G$ is the total longitudinal momentum of the system. Equation \rfn{SECOND-EQUATION-1} holds in general for the Bjorken expansion. It follows directly from the conservation law in the form $\partial_\mu T^{\mu\nu} = 0$.

\begin{figure}[t!]
\includegraphics[angle=0,width=0.9 \textwidth]{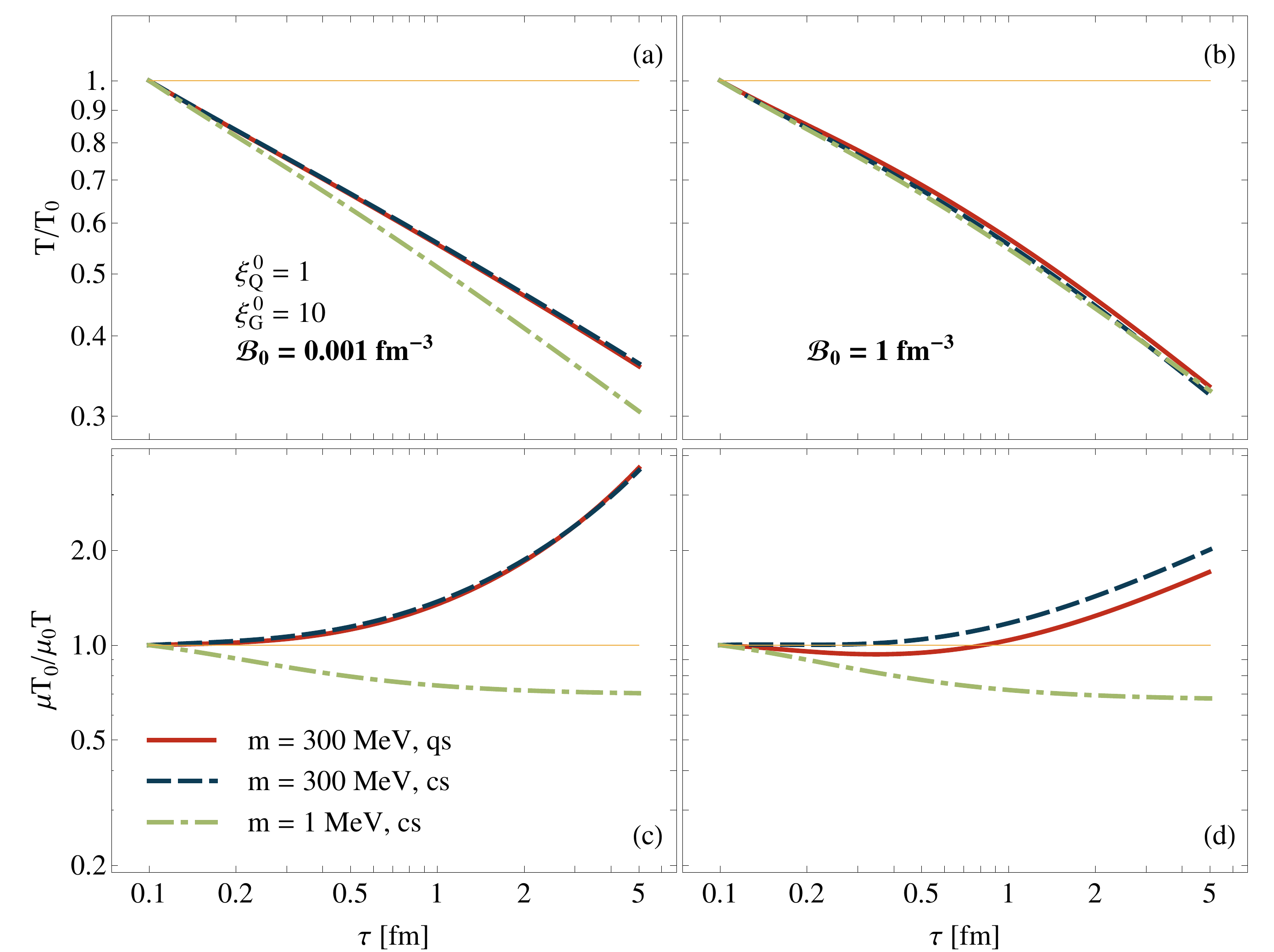} 
\caption{(Color online) Effective temperature $T$ (upper panels) and $\mu/T$ ratio (lower panels), shown as functions of the proper time $\tau$ and normalized to unity at the initial proper time $\tau=\tau_0$. Results correspond to the initial oblate-oblate configuration with the anisotropy parameters given in the figure. Three different types of lines correspond to three different choices of the statistics and the quark mass (the label ``cs'' denotes classical statistics used for both quarks and gluons, while the label ``qs'' denotes the use of Fermi-Dirac and Bose-Einstein statistics for quarks and gluons, respectively). Other parameters of the calculations are shown in the figure and discussed in the text.} 
\label{fig:TMu_oo}
\end{figure}   
%
\begin{figure}[t!]
\includegraphics[angle=0,width=0.9\textwidth]{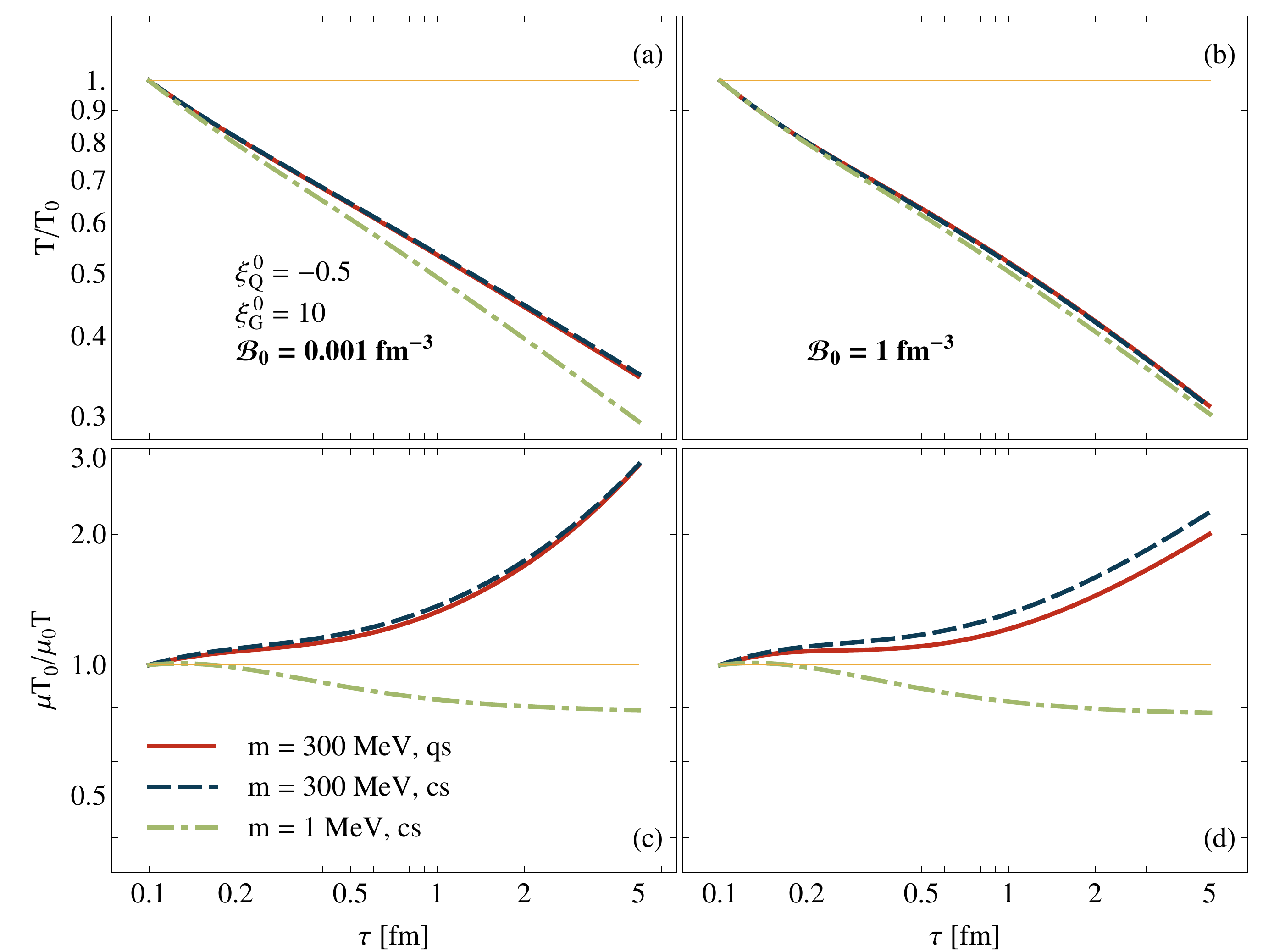}
\caption{(Color online) Same as Fig.~\ref{fig:TMu_oo} but for the initial prolate-oblate configuration with the parameters given in the figure.}
\label{fig:TMu_op}
\end{figure}
%
\begin{figure}[t!]
\includegraphics[angle=0,width=0.9\textwidth]{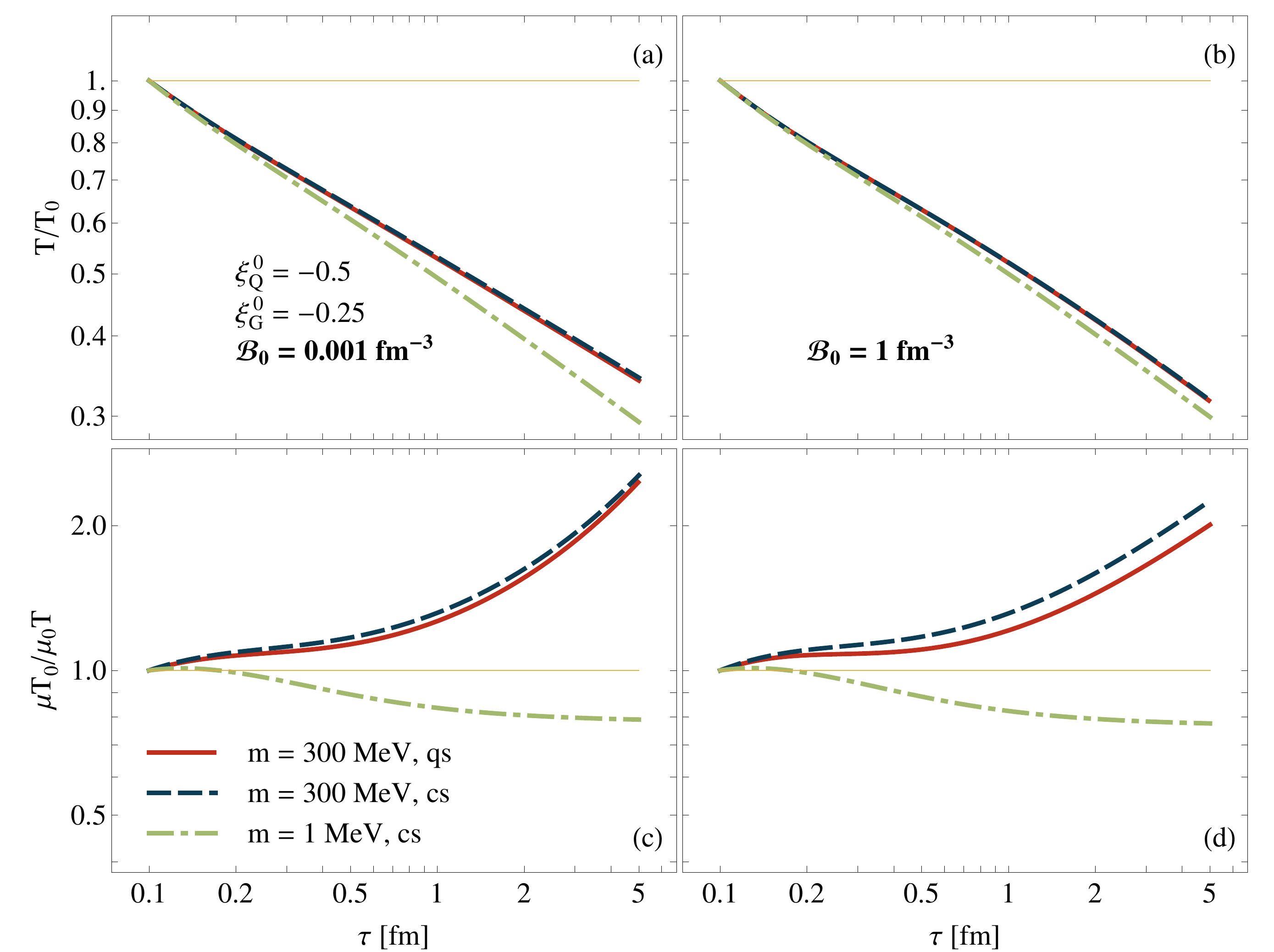} 
\caption{(Color online) Same as Figs.~\ref{fig:TMu_oo} and \ref{fig:TMu_op} but for the initial prolate-prolate configuration with the parameters given in the figure. }
\label{fig:TMu_pp}
\end{figure}
%

\section{Results}
\label{sect:res}

In this section we present the results of our numerical calculations. In all studied cases
we use a constant equilibration time
$\tau_{\rm eq}=0.25$~fm, which is the same for quark and gluon components.~\footnote{Here the main motivation comes from saving the computational time.
A  popular case used in conformal theories,  where $\teq$ is inversly proportional to the effective temperature $T$, leads to much longer
calculations due to additional integral in  Eq.~(\ref{Damp1}).} The starting proper time is $\tau_0=0.1$~fm and the evolution continues till $\tau_f=5.0$~fm (or $\tau_f=10$~fm in several cases). The initial transverse momentum scales of quarks and gluons are taken identical and always fixed to $\Lambda_{\Q}^{0}=\Lambda_{\G}^{0}=1$~GeV. The initial non-equilibrium  chemical potential $\lambda_0$ is chosen in such a way that the initial baryon number density is either ${\cal B}_0~=$~0.001~fm$^{-3}$ or ${\cal B}_0=$~1~fm$^{-3}$, see Eq.~(\ref{BApp}). 

Other initial conditions correspond to different values of the anisotropy parameters. We use three sets of the values for $\xi_{\Q}^{0}$ and $\xi_{\G}^{0}$: i) $\xi_{\Q}^{0}=1$ and $\xi_{\G}^{0}=10$, ii)  $\xi_{\Q}^{0}=-0.5$ and $\xi_{\G}^{0}=10$, and iii)  $\xi_{\Q}^{0}=-0.5$ and $\xi_{\G}^{0}=-0.25$. They correspond to oblate-oblate, prolate-oblate, and prolate-prolate initial momentum distributions of quarks and gluons, respectively. Such initial values for $\xi_{\Q}^{0}$ and $\xi_{\G}^{0}$ were used previously in Ref.~\cite{Florkowski:2015cba}. We note that different values of $\xi_{\Q}^{0}$, $\xi_{\G}^{0}$, and $\lambda_0$ imply different initial energy and baryon number densities, hence,  due to matching conditions, also different initial values of $T_0$ and $\mu_0$. We also note that the oblate-oblate initial configuration is supported by the microscopic calculations which suggest that the initial transverse pressure is much higher than the longitudinal one~\cite{Heller:2011ju,Gelis:2013rba}. 

We perform our calculations for three different choices of the particle statistics and the quark mass: in the first case both quarks and gluons are described by the classical, Boltzmann statistics~\footnote{In this case the $\pm$ sign in \rfn{feq} is neglected and $h^{\pm}_\eq(a) = \exp(-a)$.} and the quark mass is equal to 1 MeV~\footnote{Since this value of mass is much smaller than the considered temperature values, we refer sometimes to this case as to the ``massless'' one.}, in the second case we use again the classical statistics but the quark mass is 300 MeV, finally, in the third case the quarks are described by the Fermi-Dirac statistics and have the mass of 300 MeV, while the gluons are described by the Bose-Einstein statistics. The gluon mass is always set equal to zero. The case with classical statistics, ${\cal B}_0=$~0.001~fm$^{-3}$,  and negligibly small quark mass of 1~MeV agrees well with the exact massless case studied in Ref.~\cite{Florkowski:2015cba}. This agreement is used as one of the checks of our present approach. The complete set of our initial conditions is given in the tables of Appendix
\ref{s:tables}.

We note that the values of the initial conditions used in this work are to large extent arbitrary, as we want to analyse here only general features of the solutions of~Eqs.~\rfn{kineq}. With more specific systems in mind, one can choose other values of the initial parameters. 

\begin{figure}[t]
\includegraphics[angle=0,width=0.5\textwidth]{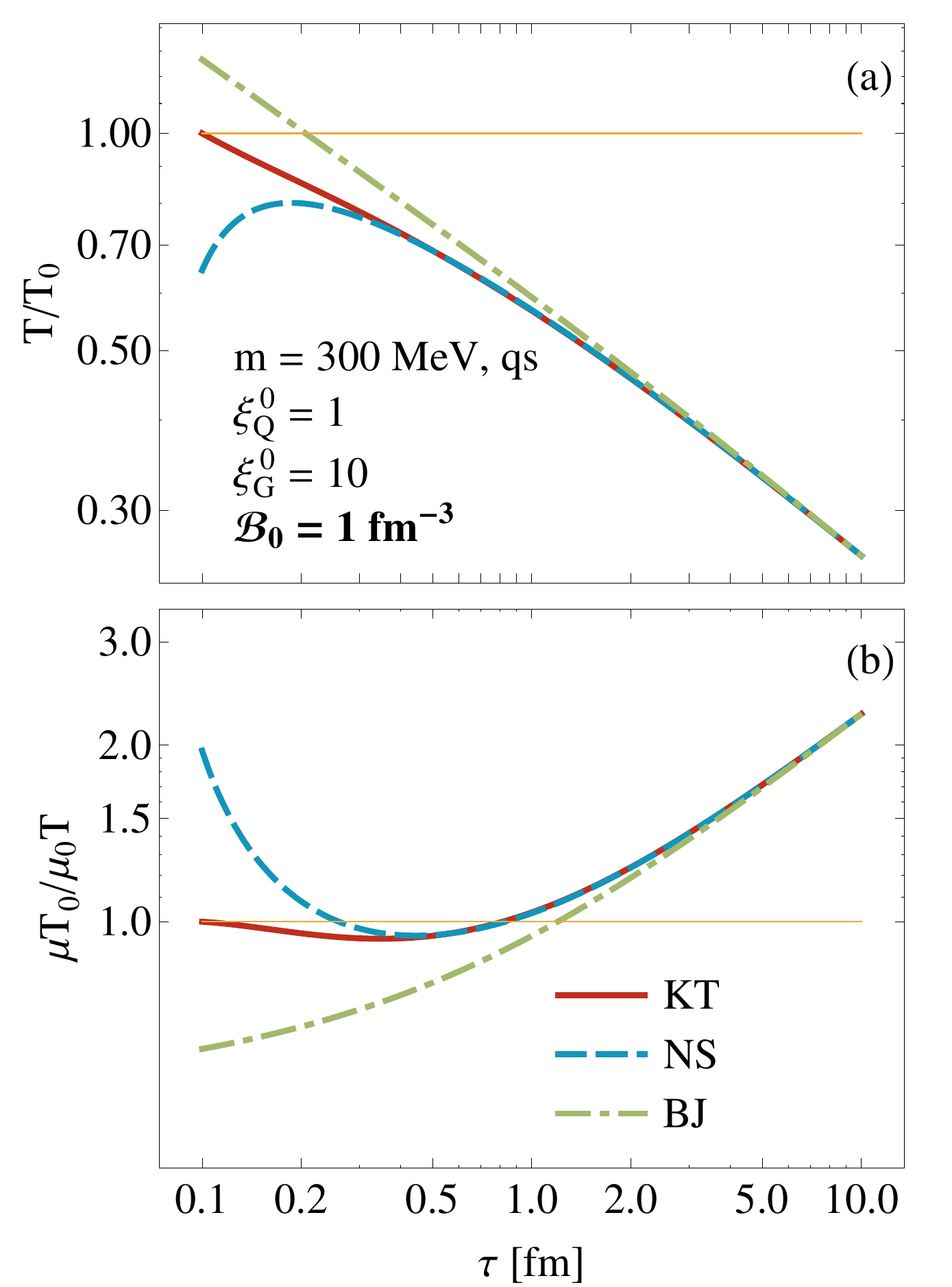}  \\
\caption{(Color online) Proper-time dependence of the ratios $T/T_0$ (a) and $\mu T_0/(\mu_0 T)$ (b) obtained in the range $\tau_0 < \tau < 10$~fm from: kinetic theory (red solid lines), perfect-fluid hydrodynamics (green dot-dashed lines), and Navier-Stokes hydrodynamics (navy blue dashed lines). All results are normalized to the initial values $T_0$ and $\mu_0$ used in the kinetic theory. The initial values of temperature and chemical potential in the hydrodynamic calculations are chosen in such a way that the final values of $T$ and $\mu$ agree with the values found in the kinetic-theory calculation. The calculations are done for the oblate-oblate initial conditions with a finite quark mass of 300 MeV, quantum statistics, and ${\cal B}_0=$~1~fm$^{-3}$.}
\label{fig:MuT_10fm_NS}
\end{figure} 
%
\begin{figure}[t!]
\includegraphics[angle=0,width=0.6\textwidth]{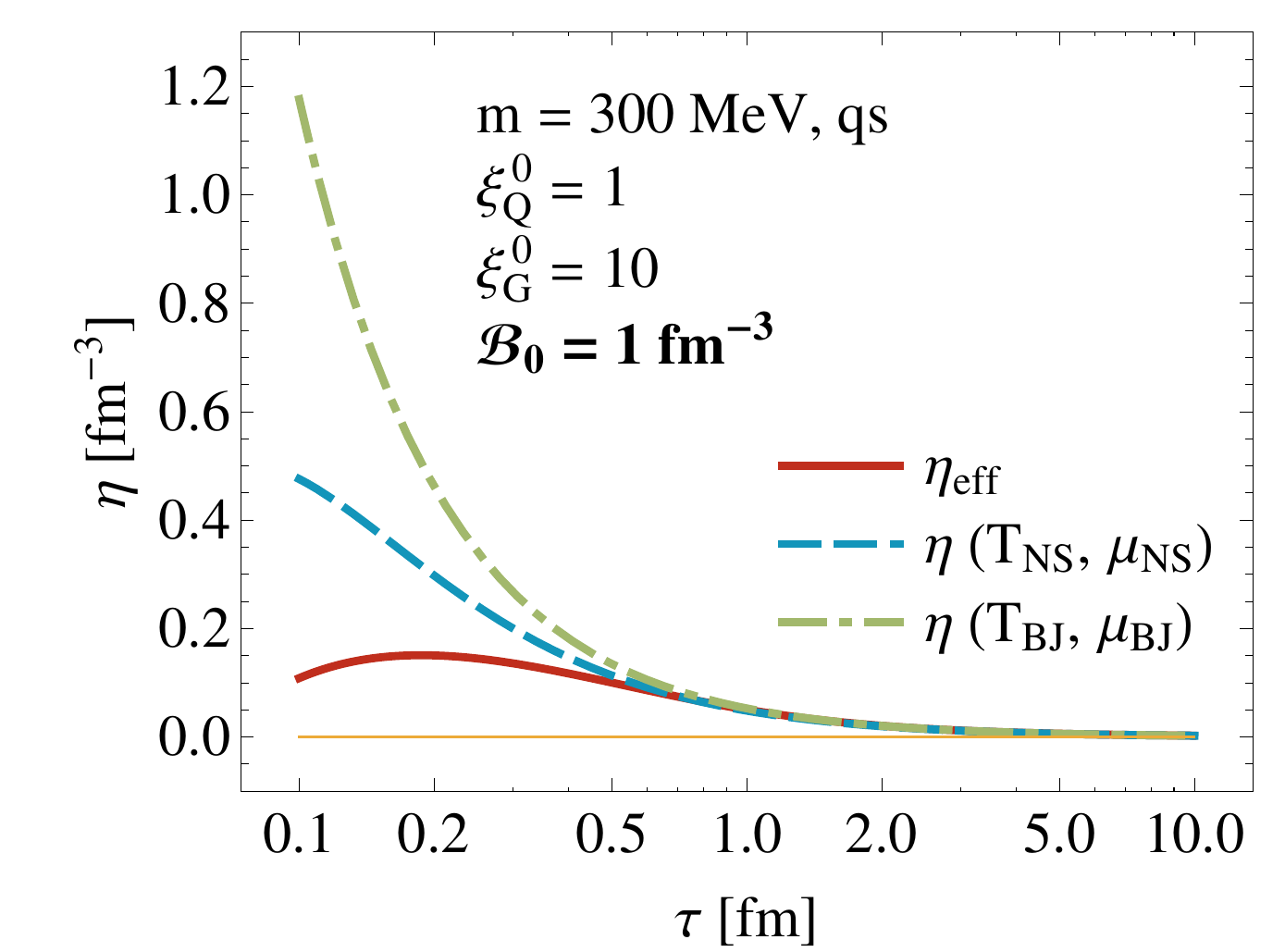} 
\caption{(Color online) Effective shear viscosity $\eta_{\rm eff}$ defined by \rf{pi1} (red solid line) and the shear viscosity coefficients $\eta$ calculated using \rf{eta} for the two $T(\tau)$ and $\mu(\tau)$ profiles, found from the perfect-fluid hydrodynamics (green dot-dashed line) and from the Navier-Stokes equations (navy blue dashed line).  The effective shear viscosity agrees well with the standard definition of $\eta$ for $\tau >  0.5$~fm. The initial conditions are the same as in Fig.~\ref{fig:MuT_10fm_NS}.}
\label{fig:MuT_10fm_Eta}
\end{figure}
%

%
\begin{figure}[t!]
\includegraphics[angle=0,width=0.495 \textwidth]{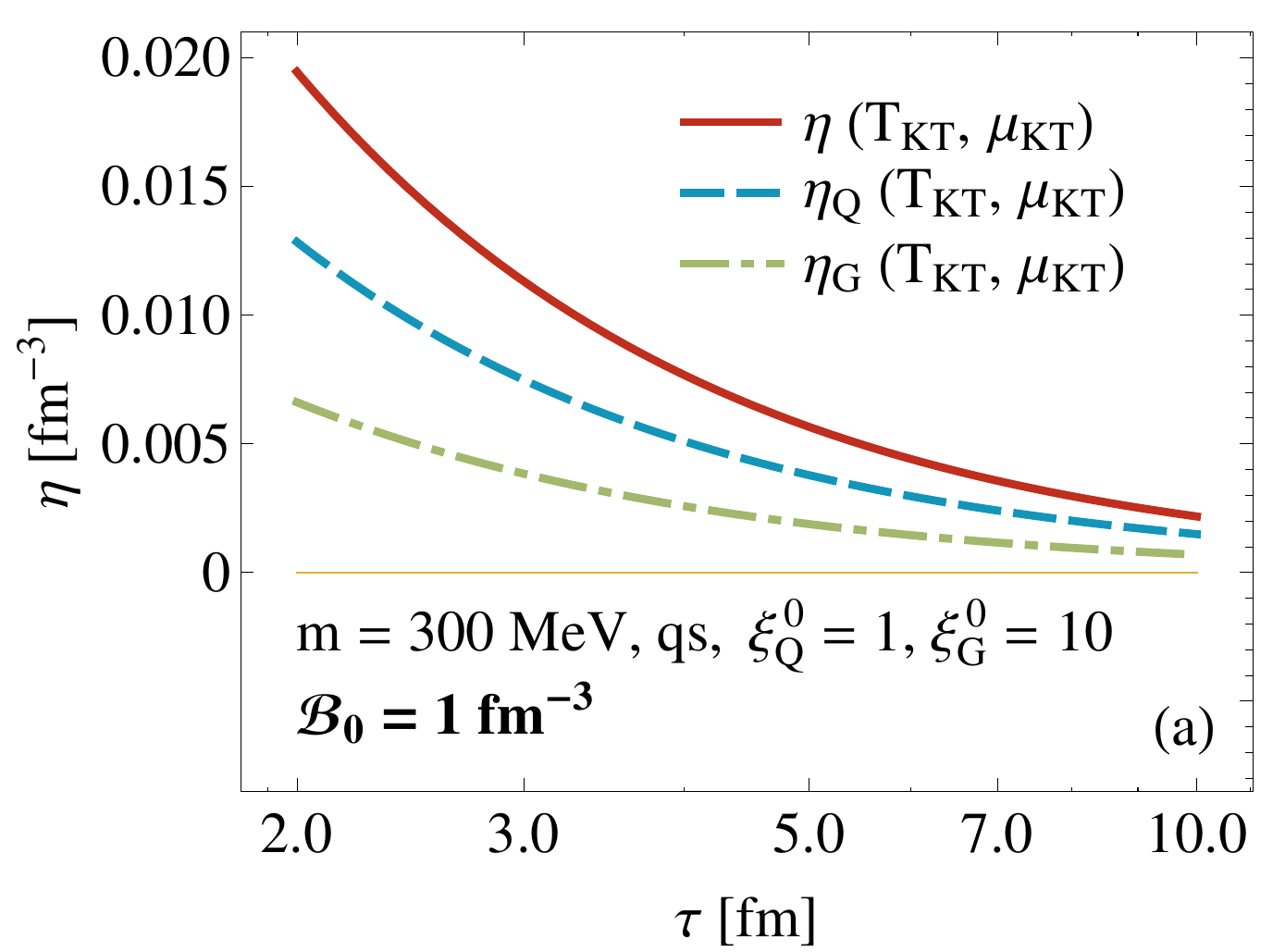} 
\includegraphics[angle=0,width=0.43 \textwidth]{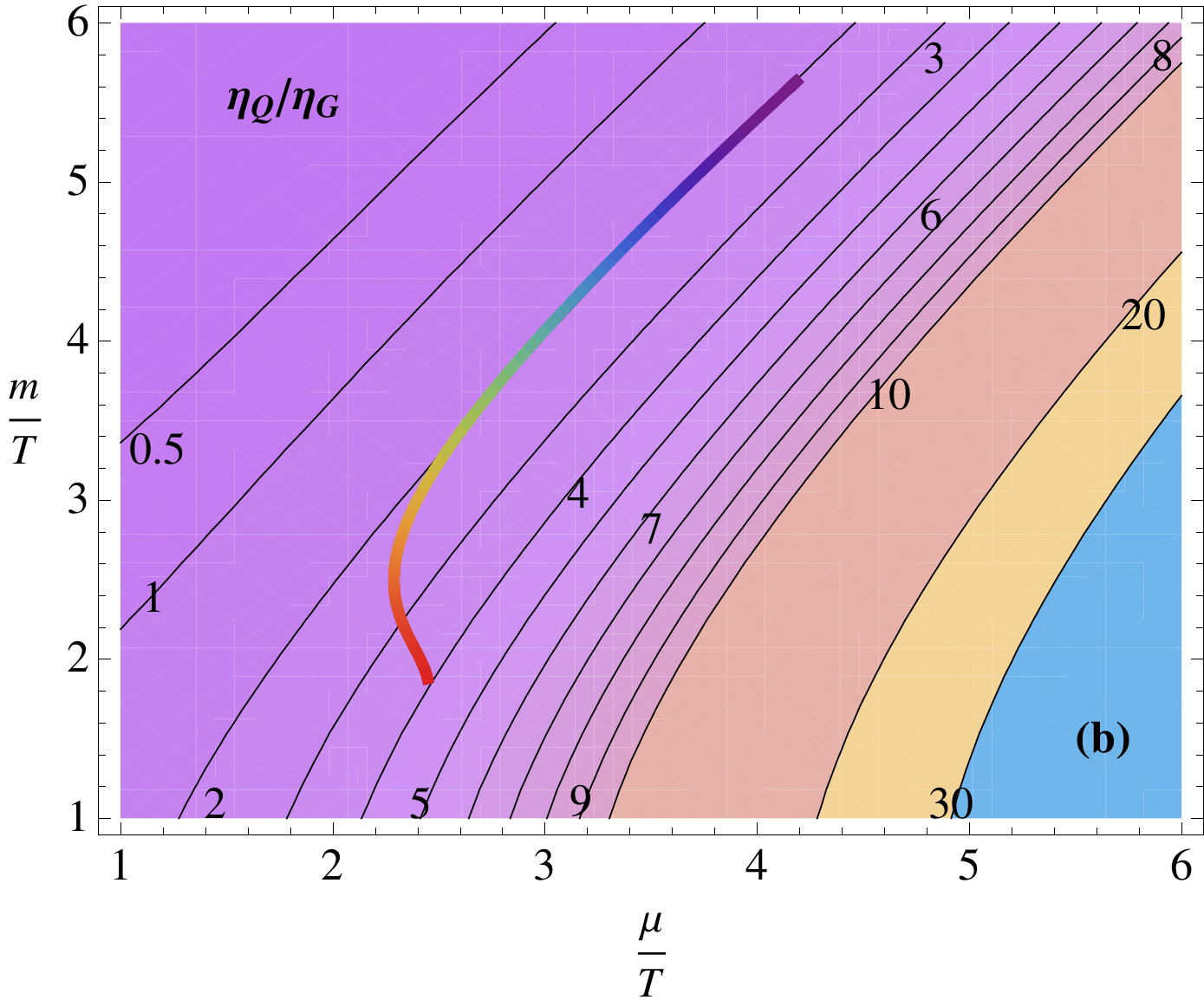} 
\caption{(Color online) Panel (a): Proper-time dependence of the shear viscosity of the mixture (red solid line), of the quark component (navy blue dashed line), and of the gluon component (green dot-dashed line), see Eqs.~\rfn{eta}, \rfn{etaQ} and \rfn{etaG}. The initial conditions are the same as in Fig.~\ref{fig:MuT_10fm_NS}. Panel (b): The ratio $\eta_\Q /\eta_G$ as a function of $m/T$ and $\mu/T$. The colored line represents the system evolution trajectory with the parameters corresponding to the panel (a).}
\label{fig:MuT_10fm_EtaQG}
\end{figure}
%

\subsection{Proper-time dependence of $T$ and $\mu/T$}
\label{sect:resTmu}

Figures \ref{fig:TMu_oo}, \ref{fig:TMu_op} and \ref{fig:TMu_pp} show the proper-time dependence of the effective temperature $T$ and $\mu/T$ ratio, which are normalised to unity at the initial time $\tau=\tau_0$. The two upper panels, (a) and (b), show temperature profiles, while the two lower panels, (c) and (d), show $\mu/T$. The two left panels, (a) and (c), correspond to the case ${\cal B}_0=$~0.001~fm$^{-3}$, and the two right panels, (b) and (d), describe the case ${\cal B}_0=$~1~fm$^{-3}$. The three figures correspond to three different initial conditions specified by the initial anisotropy parameters. Figures~ \ref{fig:TMu_oo}, \ref{fig:TMu_op}, and \ref{fig:TMu_pp} illustrate the effects of the finite mass and quantum statistics on the time evolution of $T$ and $\mu/T$. We observe that the inclusion of the finite mass (for either classical or quantum statistics) has an important effect on the $\mu/T$ ratio. For $m=300$~MeV it asymptotically increases with time, while in the $m=1$~MeV case it approaches a constant, which is expected for the massless system in the Bjorken model assuming local equilibrium. 
The finite mass has a small effect on the time dependence of the effective temperature. The latter decreases more slowly in the massive cases (especially in the ${\cal B}_0=$~0.001~fm$^{-3}$ case). The effects of quantum statistics are most visible in the $\mu/T$ proper-time dependence. 

To analyze the proper-time dependence of $T$ and $\mu/T$ in more detail, in Fig.~\ref{fig:MuT_10fm_NS} we compare the kinetic-theory (KT) results for the quantum, massive, and oblate-oblate case with hydrodynamic calculations. The latter are performed for the Bjorken perfect-fluid (BJ) and Navier-Stokes (NS) versions, see Appendix~\ref{s:NS} for definitions of these frameworks. The initial values of temperature and chemical potential in the hydrodynamic calculations are chosen in such a way that the final values of $T$ and $\mu$ agree with the values found in the kinetic-theory calculation. Although such matching is required only for the last moment of the time evolution, we see that the hydrodynamic calculations approximate very well the kinetic-theory results within a few last fermis of the time evolution. As expected, we see that the Navier-Stokes approach reproduces better the exact  kinetic-theory result, compared to the perfect-fluid calculation, as it accounts for the dissipative effects in the system.

\begin{figure}[t!]
\includegraphics[angle=0,width=0.55\textwidth]{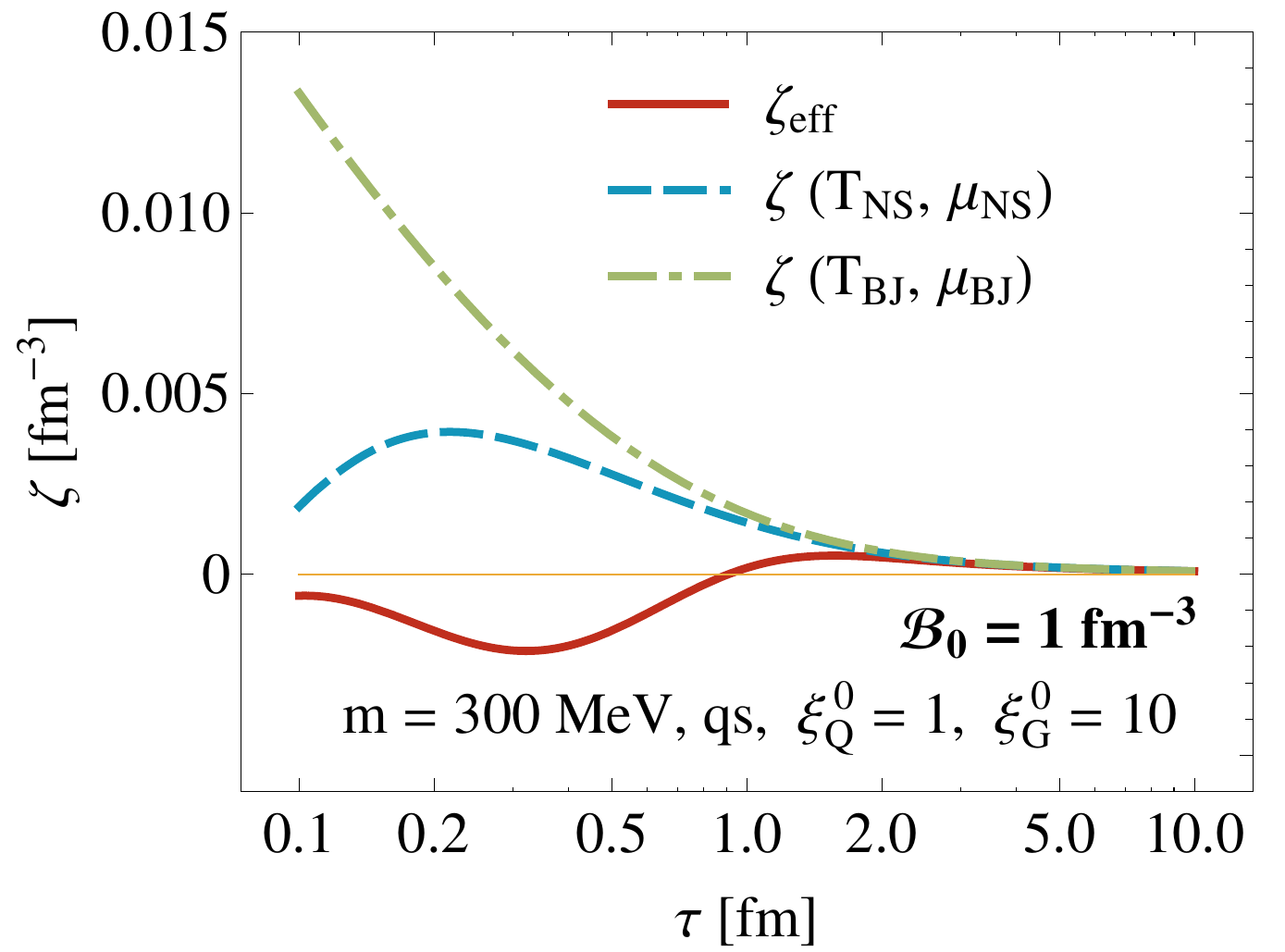}  \\
\caption{(Color online) Effective bulk viscosity $\zeta_{\rm eff}$ defined by \rf{Pi1} (red line) and the bulk viscosity coefficient $\zeta$ calculated with the help of \rf{zeta} for the two $T(\tau)$ and $\mu(\tau)$ profiles, found from the perfect-fluid hydrodynamics (green dot-dashed line) and from the Navier-Stokes equations (blue dashed line).  We find that the effective bulk viscosity agrees with the standard definition of $\zeta$ for $\tau > 2$~fm. The initial conditions are the same as in Fig.~\ref{fig:MuT_10fm_NS}. }
\label{fig:MuT_10fm_Zeta}
\end{figure}
%
\begin{figure}[t!]
\includegraphics[angle=0,width=0.55\textwidth]{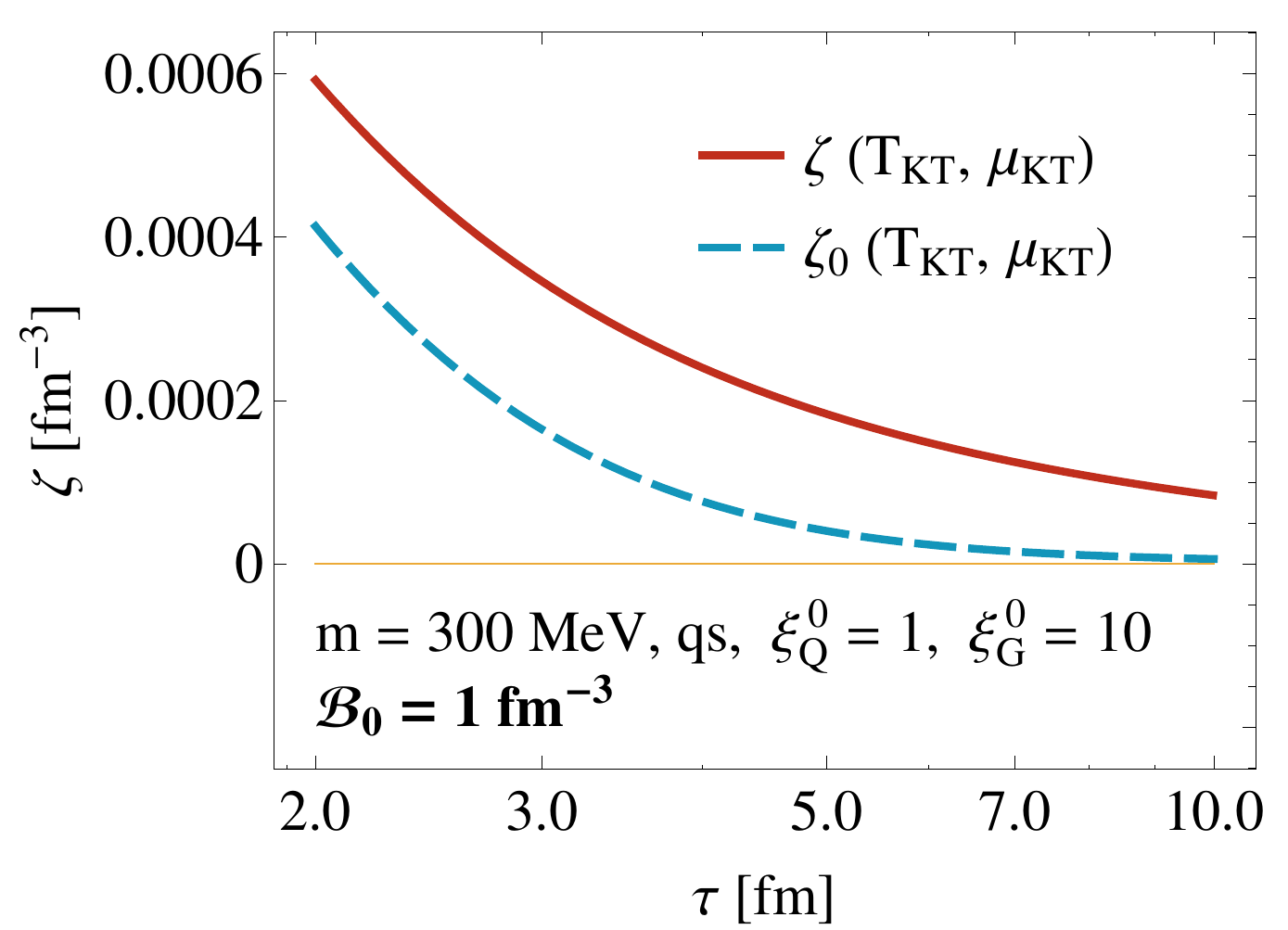}  \\
\caption{(Color online) Bulk viscosity coefficient $\zeta$ calculated with the help of \rf{zeta}  with the thermodynamic coefficients $\kappa_1$ and $\kappa_2$ determined for the whole quark-gluon system (red solid line) and the coefficient $\zeta_0$ obtained from  \rf{zeta} with $\kappa_1$ and $\kappa_2$ determined only for the quark component (blue dashed line). The initial conditions are the same as in Fig.~\ref{fig:MuT_10fm_NS}. }
\label{fig:MuT_10fm_Zeta0}
\end{figure}
%

\subsection{Hydrodynamization}
\label{sect:hydrodynamization}

\subsubsection{Shear sector}

The results shown in Fig.~\ref{fig:MuT_10fm_NS} suggest that the non-equilibrium dynamics of the system enters
rather fast the hydrodynamic regime described by the NS equations (at the stage where deviations from local
equilibrium are still substantial). Such a phenomenon was identified first in the context
of AdS/CFT calculations \cite{Heller:2011ju} and is known now as the \textit{hydrodynamization} process. To illustrate this
behaviour  in our case, we show in Fig.~\ref{fig:MuT_10fm_Eta} the proper-time dependence of the effective shear viscosity coefficient
$\eta_{\rm eff}$ defined by the expression~\cite{Florkowski:2013lya}~\footnote{We use the notation where calligraphic symbols such as ${\cal E}$, ${\cal P}_T $ or ${\cal P}_L$ refer to exact values obtained from the kinetic theory. In the situations where the system is close to equilibrium and described by the Navier-Stokes hydrodynamics we add the subscript $NS$. The standard kinetic coefficients describe the systems close to equilibrium, hence, the shear viscosity
is defined by the formula $\eta = \f{\tau}{2}  \left({\cal P}_T - {\cal P}_L \right)_{\rm NS}$ and the bulk viscosity by $\zeta = - \tau \Pi_{\rm NS}$, see App.~\ref{s:NS}. If we use the exact  kinetic-theory values
on the right-hand sides of these definitions we deal with effective values, which should agree with the standard definitions for systems being close to local equilibrium. }

\bel{pi1}
\eta_{\rm eff} = \f{\tau}{2}  \left({\cal P}_T - {\cal P}_L \right) ,
\eel
see Eqs.~\ref{PLPTns}.
The \textit{effective shear viscosity} (solid red line in Fig.~\ref{fig:MuT_10fm_Eta}) is compared with the standard shear viscosity coefficient, $\eta$, valid for the system close to equilibrium. For the quark-gluon mixture
the latter is defined as the sum of the quark and gluon coefficients,~\footnote{For general collision kernels, the total shear viscosity (although written formally as a sum of the individual contributions) may  not be a simple sum of {\it independent} terms, for example, see~\cite{Itakura:2007mx}.}
\bel{eta}
\eta = \eta_\Q + \eta_\G,
\eel
where following~\cite{Sasaki:2008fg}, see also \cite{Bozek:2009dw,Chakraborty:2010fr,Bluhm:2010qf}, we use
\bel{etaQ}
\eta_\Q = \f{g_\Q \teq}{15 T} \int_0^\infty \f{dp\, p^6}{2 \pi^2 (m^2+ p^2)} 
\left[ f_{\Qp, \eq} \left(1-f_{\Qp, \eq} \right) + f_{\Qm, \eq} \left(1-f_{\Qm, \eq} \right) \right],
\eel
\bel{etaG}
\eta_\G = \f{g_\G \teq}{15 T} \int_0^\infty \f{dp\, p^4}{2 \pi^2} f_{\G, \eq} \left(1+f_{\G, \eq} \right).
\eel
The coefficient $\eta$ is calculated as a function of  $T$ and $\mu$ obtained either from the perfect-fluid  
(green dot-dashed line in Fig.~\ref{fig:MuT_10fm_Eta}) or  NS hydrodynamic calculation 
(navy blue dashed line in Fig.~\ref{fig:MuT_10fm_Eta}). In the two cases 
we find that $\eta_{\rm eff}$ agrees very well with $\eta$ for $\tau > 0.5$~fm which is about two times the 
relaxation time. Thus, in the shear sector we observe a very fast approach to the hydrodynamic NS regime. It is important to notice that 
the agreement with the NS description is reached when $\eta$ is significantly different from zero, which supports the idea 
that the hydrodynamic description becomes appropriate before the system thermalises, i.e., before the state of local thermal 
equilibrium with ${\cal P}_T \approx {\cal P}_L$ is reached.

In panel (a) in Fig.~\ref{fig:MuT_10fm_EtaQG} we show the proper-time dependence of the shear viscosity of the mixture (red solid line)
and compare it with the shear viscosity of the quark component (navy blue dashed line) and the gluon component (green dot-dashed line), see Eqs.~\rfn{eta}, \rfn{etaQ} and \rfn{etaG}. The initial conditions are the same as in Fig.~\ref{fig:MuT_10fm_NS}. The results shown in Fig.~\ref{fig:MuT_10fm_EtaQG}
show that the shear viscosity of the mixture is dominated by the shear viscosity of quarks thoughout the system evolution. The information complementary to panel (a) is provided in the panel (b) in Fig.~\ref{fig:MuT_10fm_EtaQG} where we present the ratio  $\eta_\Q/\eta_G$ as a function of $m/T$ and $\mu/T$ (contour lines) together with the system trajectory (colored line). 

\begin{figure}[t]
\includegraphics[angle=0,width=0.9\textwidth]{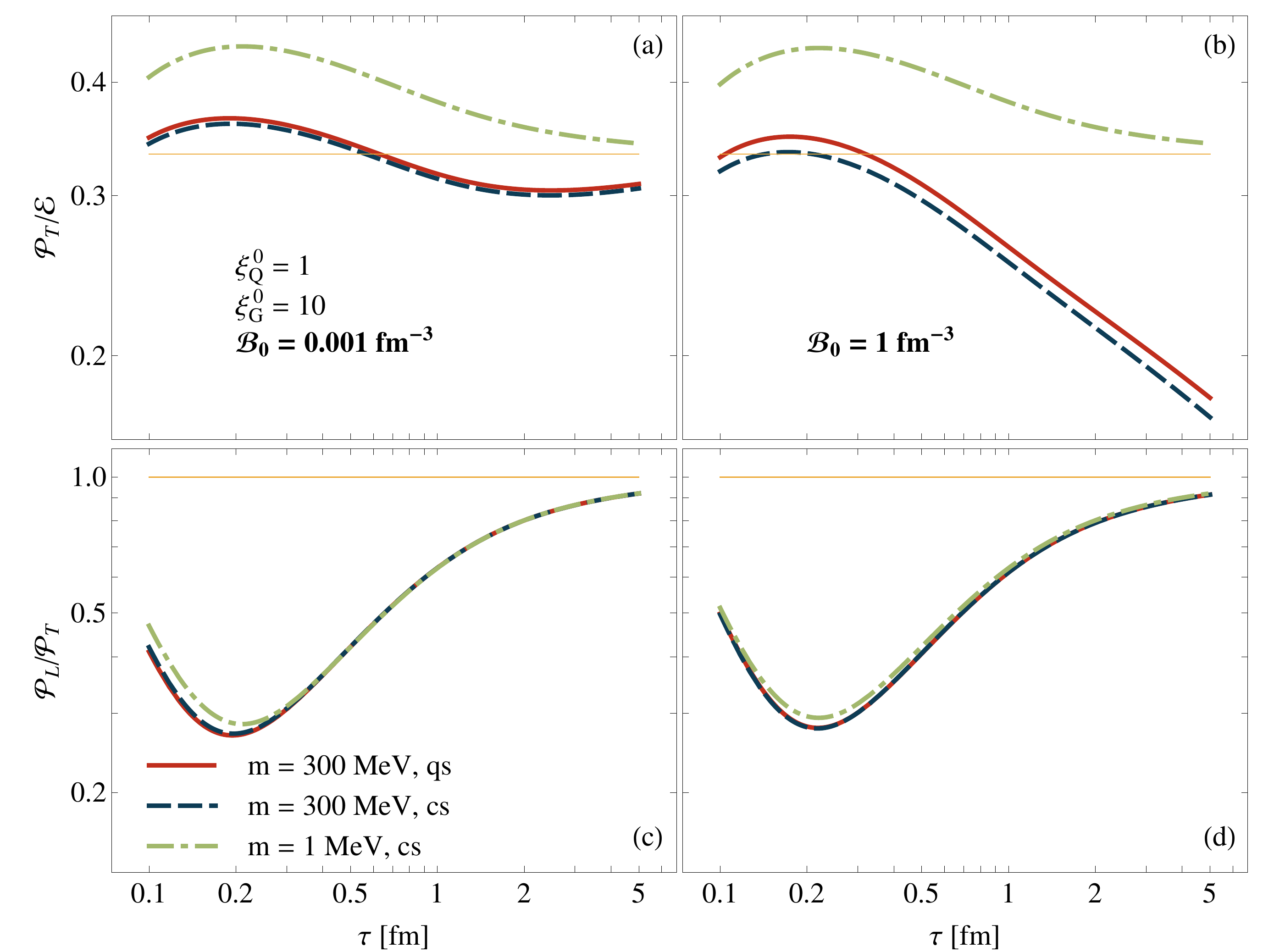} 
\caption{(Color online) ${\cal P}_T/{\cal E}$ (upper panels) and ${\cal P}_L/{\cal P}_T$ (lower panels) for initially oblate-oblate system. Green solid lines correspond to ``massless'' quarks and classical distribution functions, black dashed lines to the massive quarks and classical distribution functions, while red solid lines are for massive quarks and quantum distributions. Left (right) panels describe the results for ${\cal B}_0=$~0.001~fm$^{-3}$ (${\cal B}_0=$~1~fm$^{-3}$).
 }
\label{fig:ooPP}
\end{figure}
%
\begin{figure}[t]
\includegraphics[angle=0,width=0.9\textwidth]{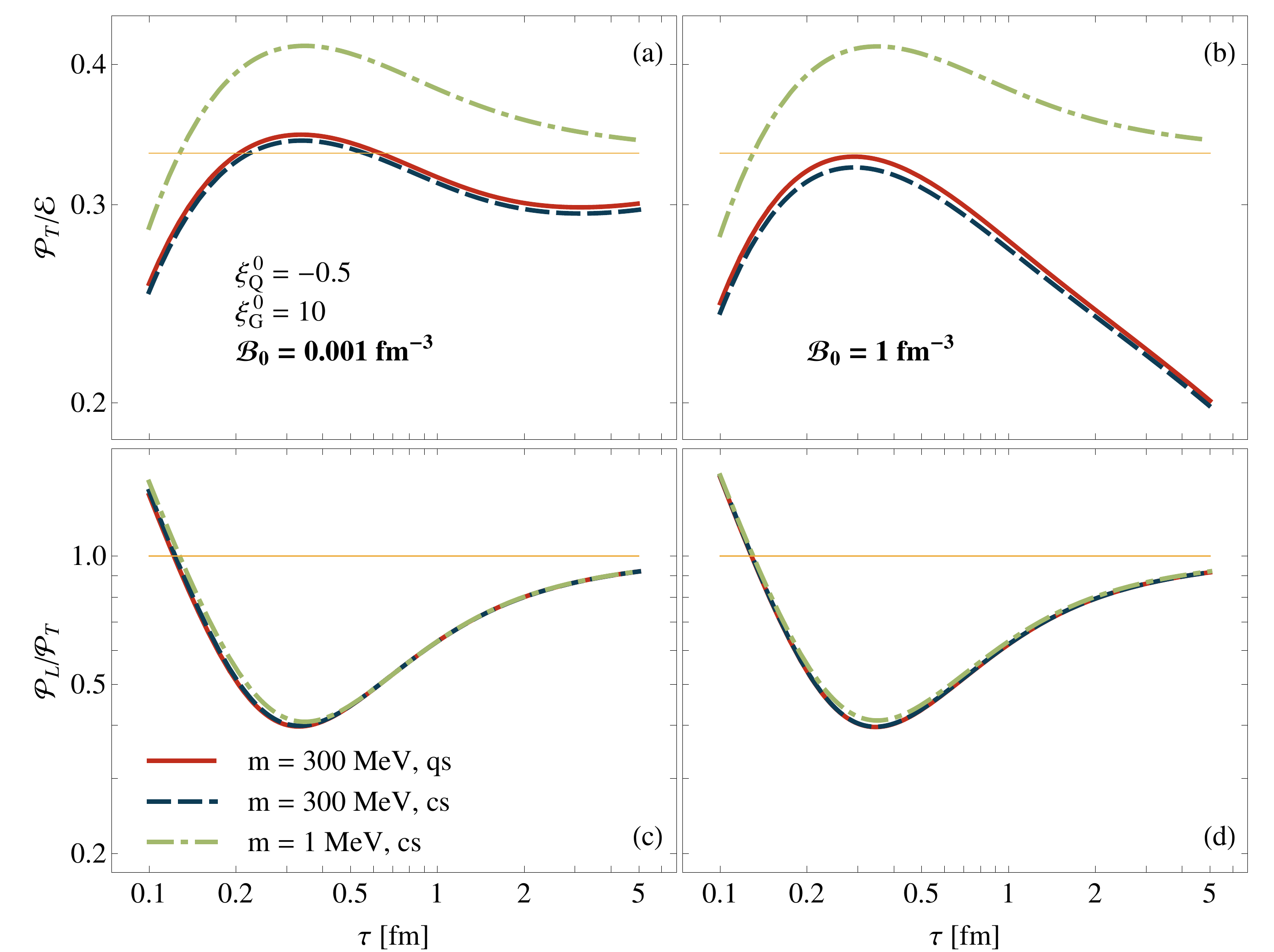} 
\caption{(Color online) Same as Fig. \ref{fig:ooPP} but for initially prolate-oblate system.}
\label{fig:opPP}
\end{figure} 
%
\begin{figure}[t]
\includegraphics[angle=0,width=0.9\textwidth]{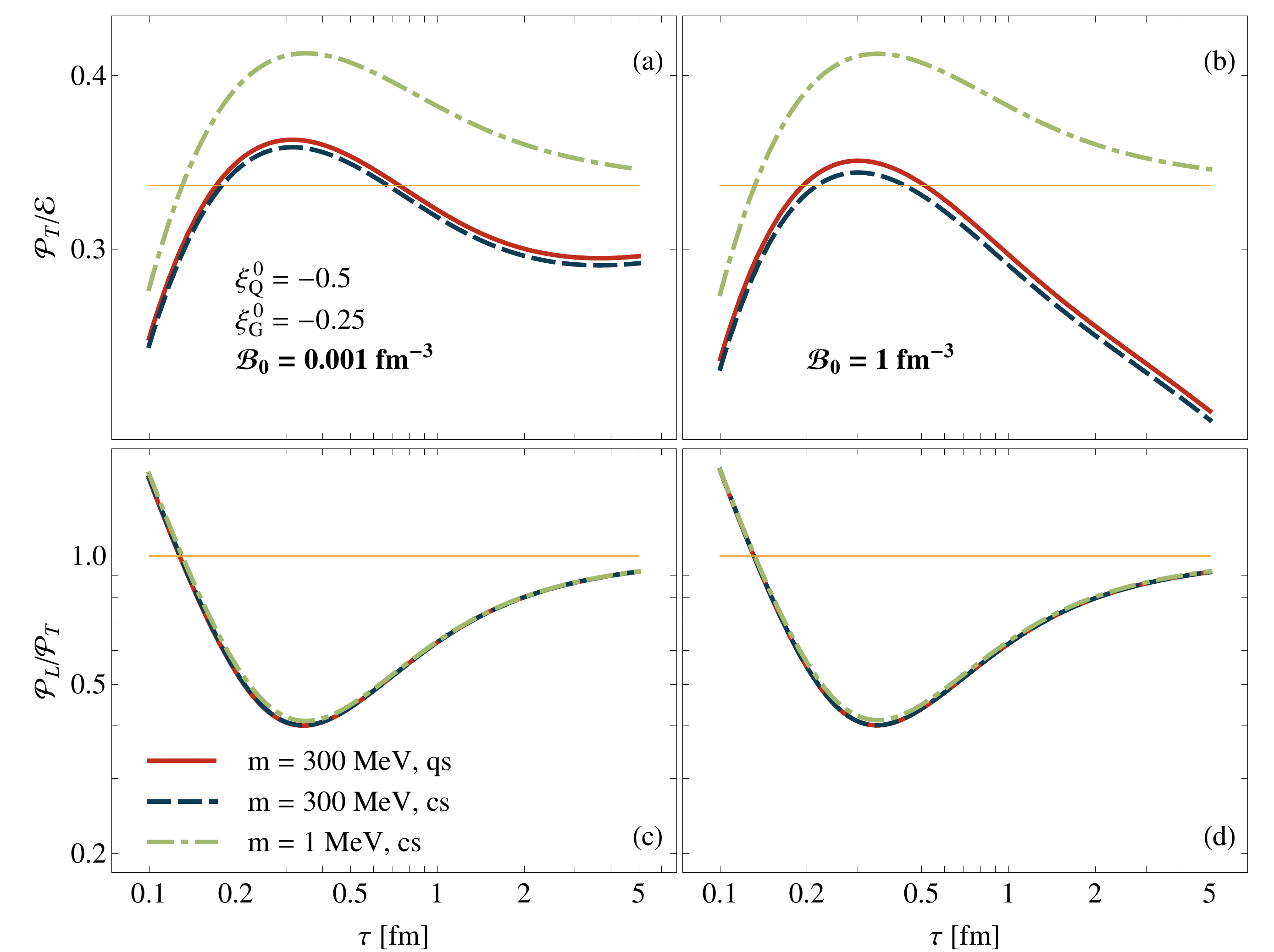} 
\caption{(Color online) Same as Fig. \ref{fig:ooPP} but for initially prolate-prolate system. }
\label{fig:ppPP}
\end{figure}  

\subsubsection{Bulk sector}  
Similarly to the shear-viscosity effects we can analyse the bulk sector, where we define the \textit{effective bulk viscosity} by the expression
\bel{Pi1}
\zeta_{\rm eff} = - \tau \Pi,
\eel
where $\Pi$ is the exact bulk pressure
\bel{Pi2}
\Pi = \f{1}{3} \left({\cal P}_L + 2 {\cal P}_T - 3 {\cal P}^{\rm eq}  \right).
\eel
The time dependence of the effective bulk viscosity is compared in Fig.~\ref{fig:MuT_10fm_Zeta} with the time dependence of the bulk viscosity coefficient
given by the expression
\begin{eqnarray}
\label{zeta}
\zeta &=& \f{g_\Q m^2 \teq}{3 T} \int_0^\infty \f{dp\, p^2}{2 \pi^2} 
\left[ \left( f_{\Qp, \eq} \left(1-f_{\Qp, \eq} \right) +  f_{\Qm, \eq} \left(1-f_{\Qm, \eq} \right) \right)
\left( \kappa_1 - \f{p^2}{3(m^2+p^2)} \right) \right. \nn \\
&& \hspace{3.5cm}  \left. +
\left( f_{\Qp, \eq} \left(1-f_{\Qp, \eq} \right) -  f_{\Qm, \eq} \left(1-f_{\Qm, \eq} \right) \right) 
\f{\kappa_2}{\sqrt{m^2 +p^2}} \right],
\end{eqnarray}
where $\kappa_1$ and $\kappa_2$ are defined by the thermodynamic derivatives
\bel{kappas}
\kappa_1(T,\mu) =  \left( \f{\partial {\cal P}^{\eq} }{\partial {\cal E}^{\eq}} \right)_{{\cal B}^{\eq}}, \qquad
\kappa_2(T,\mu) =  \f{1}{3} \left( \f{\partial {\cal P}^{\eq} }{\partial {\cal B}^{\eq} } \right)_{{\cal E}^{\eq}}.
\eel
For a simple fluid with zero baryon density, the coefficient $\kappa_1$ becomes equal to the sound velocity squared. 
The steps leading to \rf{zeta} are described in more detail in Appendix \ref{s:shearbulk}. The form of  \rfn{zeta} agrees with that
given in \cite{Sasaki:2008fg} for fermions. There is, however, one important difference between our approach and that of
\cite{Sasaki:2008fg}. In Ref.~\cite{Sasaki:2008fg} a simple system of fermions is considered and \rfn{zeta} includes 
the derivatives \rfn{kappas} where only fermionic thermodynamic functions appear. In our case we deal with
a mixture and we have checked that \rfn{kappas} should include the total thermodynamic functions being the
sums of quark and gluon contributions. Thus, although the bulk viscosity of a quark-gluon mixture is given by
the formula known for massive quarks (and $\zeta=0$ if $m=0$), the use of the full thermodynamic functions 
in \rfn{kappas} means that although gluons are considered  massless  they  contribute to the bulk viscosity of the full system.

Similarly as in the shear sector, we can see in Fig.~\ref{fig:MuT_10fm_Zeta}  that $\zeta_{\rm eff}(\tau)$ 
approaches $\zeta(\tau)$, however, the agreement is reached 
for significantly larger times, $\tau > 2$~fm. This means that the hydrodynamization of the bulk sector is slower and follows the 
hydrodynamization of the shear sector.  Observations that the hydrodynamization in the shear sector may happen before
the hydrodynamization in the bulk sector have been done recently in  Ref.~\cite{Attems:2017zam} within the non-conformal models using the \textit{gauge/gravity correspondence}, where the
hydrodynamization in the bulk sector has been dubbed the \textit{EoSization} process. In this scenario first ${\cal P}_L$ and ${\cal P}_T$
tend to a common value ${\bar {\cal P}} \neq {\cal P}^{\rm eq}$ and, subsequently, ${\bar {\cal P}}$ approaches ${\cal P}^{\rm eq}$, which signals establishing equation of state of the system.

To visualize the importance of the gluon degrees of freedom in expressions \rfn{kappas} for the bulk viscosity of the mixture in Fig.~\ref{fig:MuT_10fm_Zeta0} we show the bulk viscosity coefficient $\zeta$ and compare it with the coefficient $\zeta_0$
that has been calculated in the same way as $\zeta$ except that the thermodynamic coefficients $\kappa_1$ and $\kappa_2$ of the former
were calculated only for the quark component. We find that neglecting the gluon contribution in $\kappa_1$ and $\kappa_2$
changes substantially the values of $\zeta$ making it significantly smaller. This finding indicates that gluons, although, massless,
contribute to the bulk viscosity of a quark-gluon mixture. The necessary requirement for this effect is, however, that quarks
are massive.

\subsubsection{ ${\cal P}_T/{\cal E}$ and ${\cal P}_L/{\cal P}_T$ ratios}

Figures \ref{fig:ooPP}, \ref{fig:opPP}, and \ref{fig:ppPP} correspond to Figs.  \ref{fig:TMu_oo}, \ref{fig:TMu_op}, 
and \ref{fig:TMu_pp}, respectively, and show the time dependence of the ratios ${\cal P}_T/{\cal E}$ (upper panels)
and ${\cal P}_L/{\cal P}_T$ (lower panels). In the case of quarks with a very small mass (green dot-dashed lines)
the ratios ${\cal P}_T/{\cal E}$ tend to 1/3 as expected for massless systems approaching local equilibrium. The ratios 
${\cal P}_L/{\cal P}_T$ in all studied cases tend to unity which again reflects equilibration of the system. 
Interestingly, the ratios ${\cal P}_L/{\cal P}_T$ very weakly depend on the quark mass and the choice of the
statistics.

\begin{figure}[h!]
\includegraphics[angle=0,width=0.9\textwidth]{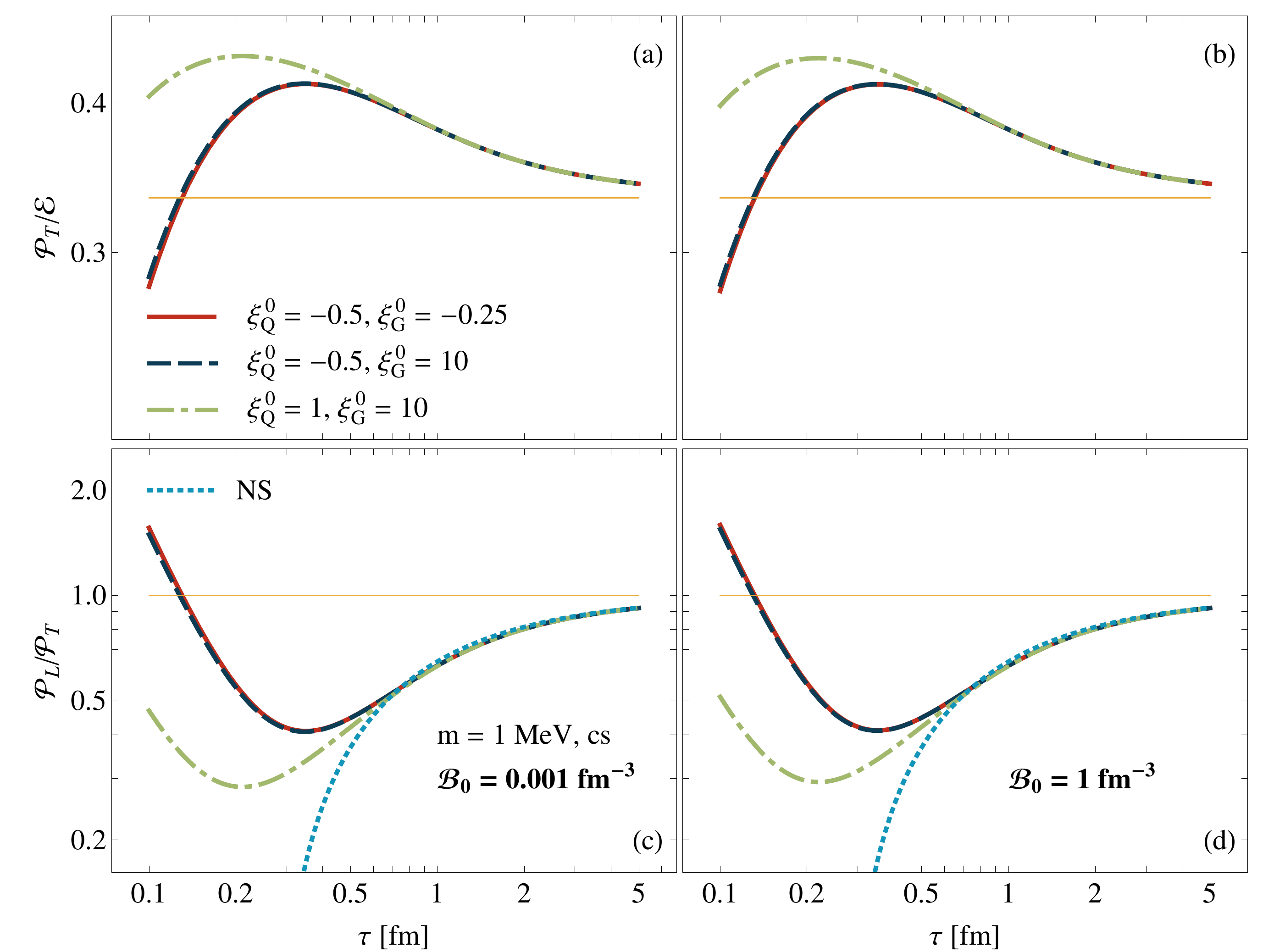} 
\caption{(Color online) ${\cal P}_T/{\cal E}$ (upper panels) and ${\cal P}_L/{\cal P}_T$ (lower  panels) for ``massless'' quarks and classical distribution functions. Green dot-dashed, navy blue dashed, and red solid lines describe the results for the oblate-oblate, prolate-oblate, and prolate-prolate initial conditions.
The blue dotted line describes $({\cal P}_L/{\cal P}_T)_{\rm NS}$ obtained from the Navier-Stokes hydrodynamics. 
}
\label{fig:m1cl_oo_op_pp}
\end{figure}
%
\begin{figure}[h!]
\includegraphics[angle=0,width=0.9\textwidth]{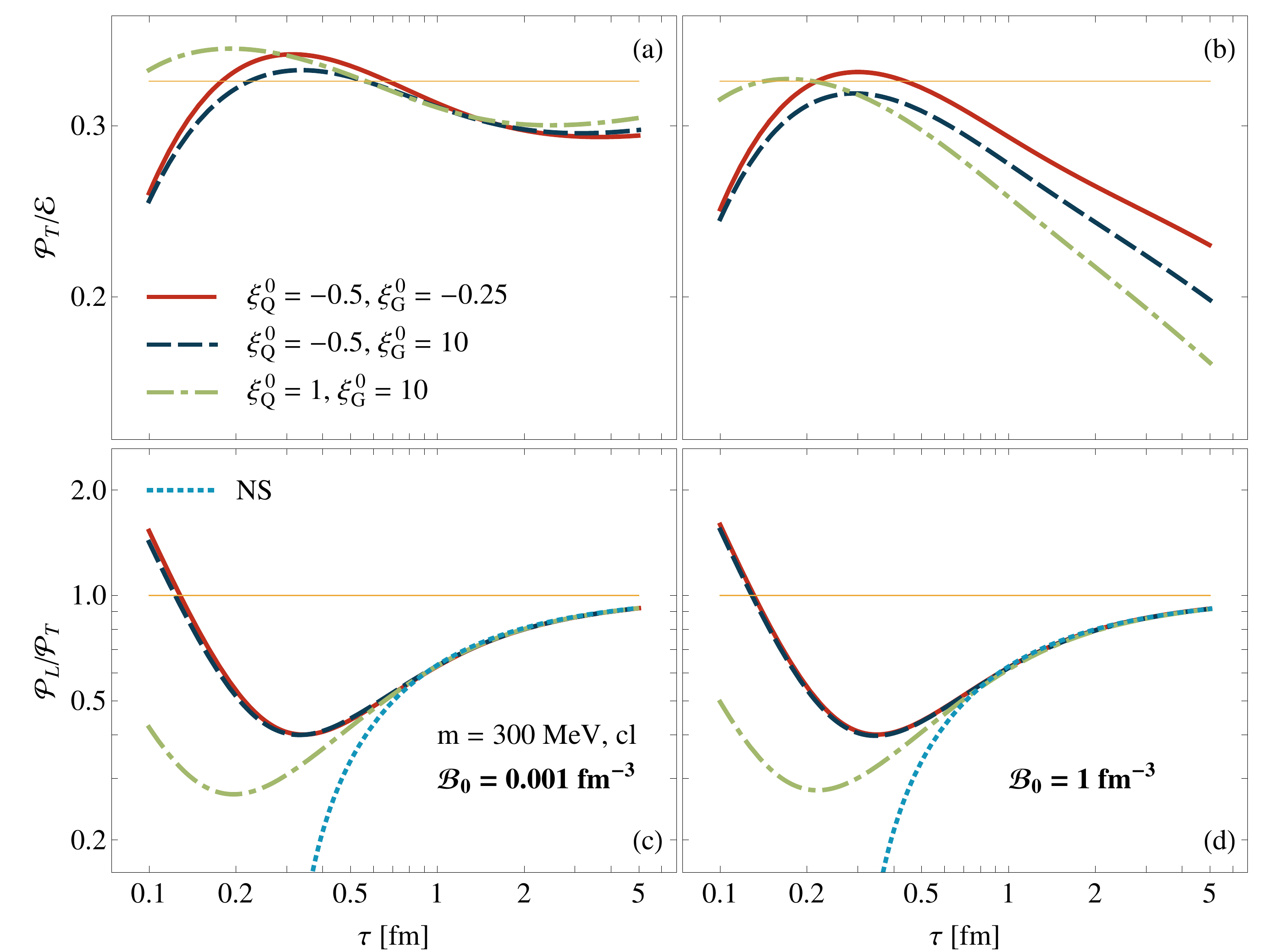} 
\caption{(Color online) Same as Fig.~\ref{fig:m1cl_oo_op_pp} but for massive quarks and classical statistics.
}
\label{fig:m300cl_oo_op_pp}
\end{figure} 
%
\begin{figure}[h!]
\includegraphics[angle=0,width=0.9\textwidth]{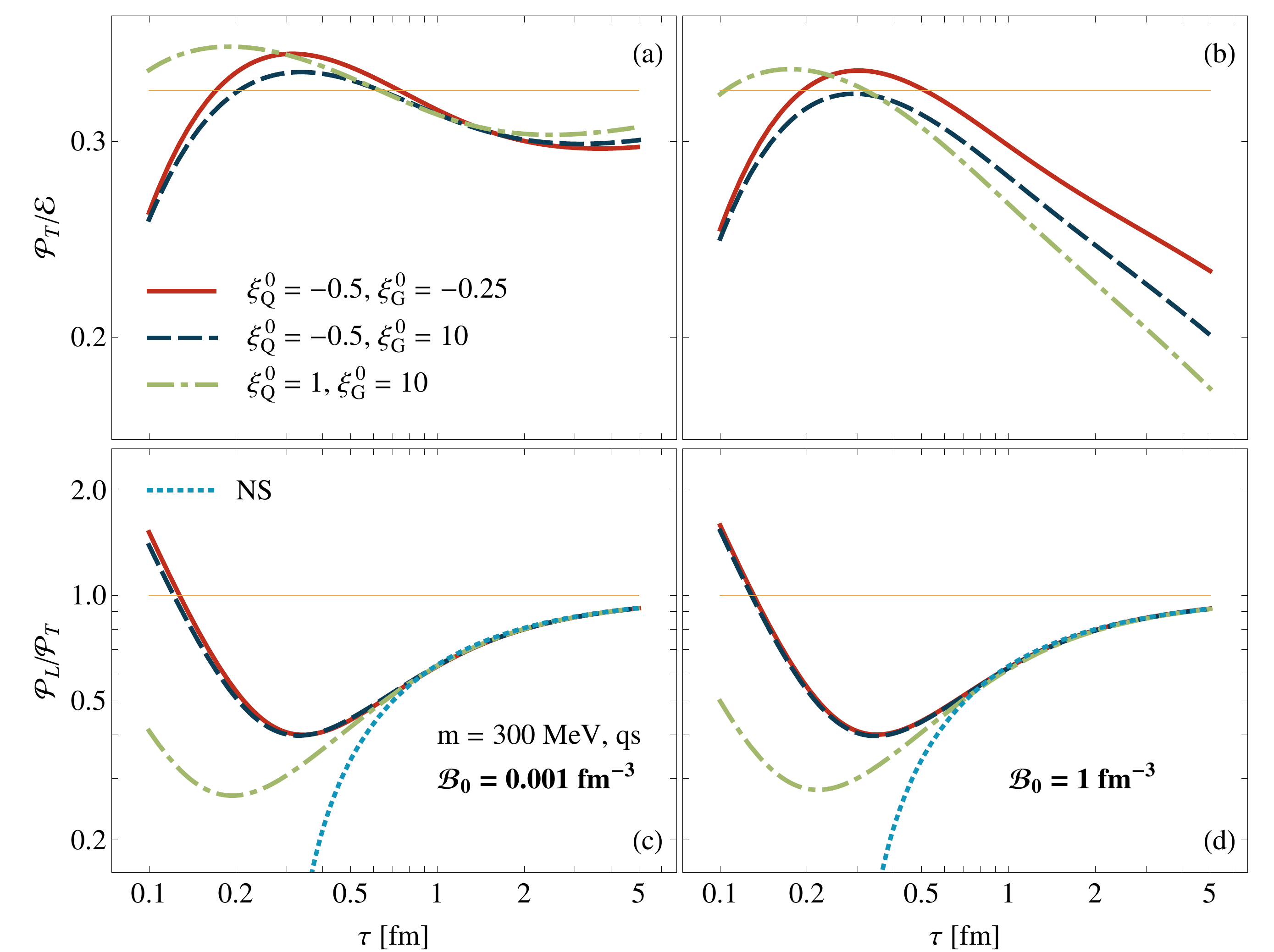} 
\caption{(Color online) Same as Fig.~\ref{fig:m1cl_oo_op_pp} but for massive quarks and quantum statistics.
}
\label{fig:m300qs_oo_op_pp}
\end{figure}

\subsection{Scaling properties}
\label{sect:scaling}

Each panel of Figs. \ref{fig:ooPP}, \ref{fig:opPP}, and \ref{fig:ppPP} shows our results obtained for different values of the quark 
mass and particle statistics but for the same initial anisotropies. In Figs. \ref{fig:m1cl_oo_op_pp}, \ref{fig:m300cl_oo_op_pp}, 
and \ref{fig:m300qs_oo_op_pp} we rearrange this information showing in each panel our results obtained for different initial 
anisotropies, i.e., for oblate-oblate, prolate-oblate, and prolate-prolate initial quark and gluon distributions. Figures~\ref{fig:m1cl_oo_op_pp}, 
\ref{fig:m300cl_oo_op_pp}, and \ref{fig:m300qs_oo_op_pp}  collect the results for different mass and statistics.
The most striking feature of our results presented in these figures is that the ${\cal P}_L/{\cal P}_T$ ratios
(shown in lower panels) converge to the same values, although they describe the system evolutions starting
from completely different initial conditions.

The origin of this behaviour can be found if we analyse the NS formula for the ${\cal P}_L/{\cal P}_T$ ratio.
Let us first consider the massless case where we may neglect the bulk viscosity and write
\bel{PLPTNS}
\left( \f{{\cal P}_L}{{\cal P}_T}  \right)_{\rm NS} =
\f{ {\cal P}^{\Q, \rm eq} -4 \eta_Q/(3 \tau)+ {\cal P}^{\G, \rm eq} -4 \eta_G/(3 \tau) }
{{\cal P}^{\Q, \rm eq} +2 \eta_Q/(3 \tau)+ {\cal P}^{\G, \rm eq} +2 \eta_G/(3 \tau)}.
\eel
Assuming in addition that the baryon number density is zero, we may use the following relations connecting 
the shear viscosity with equilibrium pressure~\footnote{See our discussion below \rf{shear4}.}: 
\bel{etaP}
\eta_Q = \f{4}{5} \teq {\cal P}^{\Q, \rm eq}, \quad \eta_G = \f{4}{5} \teq {\cal P}^{\G, \rm eq}.
\eel
It is interesting to note that the coefficient 4/5 is the same for quarks and gluons, hence
\bel{PLPTNSa}
\left( \f{{\cal P}_L}{{\cal P}_T}  \right)_{\rm NS}  =
\f{ 1 - 16 \teq/(15 \tau)}{1+ 8 \teq/(15 \tau)},
\eel
which explains the late-time dependence of ${\cal P}_L/{\cal P}_T$ on the proper time only, observed in panel (c) 
of Fig. \ref{fig:m1cl_oo_op_pp}. We note that if the relaxation time is inversely proportional to the temperature, 
\rf{PLPTNSa} indicates that  $\left({\cal P}_L/{\cal P}_T\right)_{\rm NS}$ depends on the product of $\tau$ and $T$, which is expected for 
conformal systems and related to the existence of a hydrodynamic attractor for such systems~\cite{Heller:2015dha,Romatschke:2016hle,Spalinski:2016fnj,Romatschke:2017vte,Spalinski:2017mel,Strickland:2017kux}.  It turns out that the inclusion of the finite mass and baryon chemical potential 
(with the values studied in this work) affects very little Eqs.~\rfn{etaP} connecting the shear viscosity with pressure. The main difference is that the
coefficient 4/5 is slightly changed. It should be replaced by an effective value obtained for the studied range of $T$ and~$\mu$.

\begin{figure}[t]
\begin{center}
\subfigure{\includegraphics[angle=0,width=0.47\textwidth]{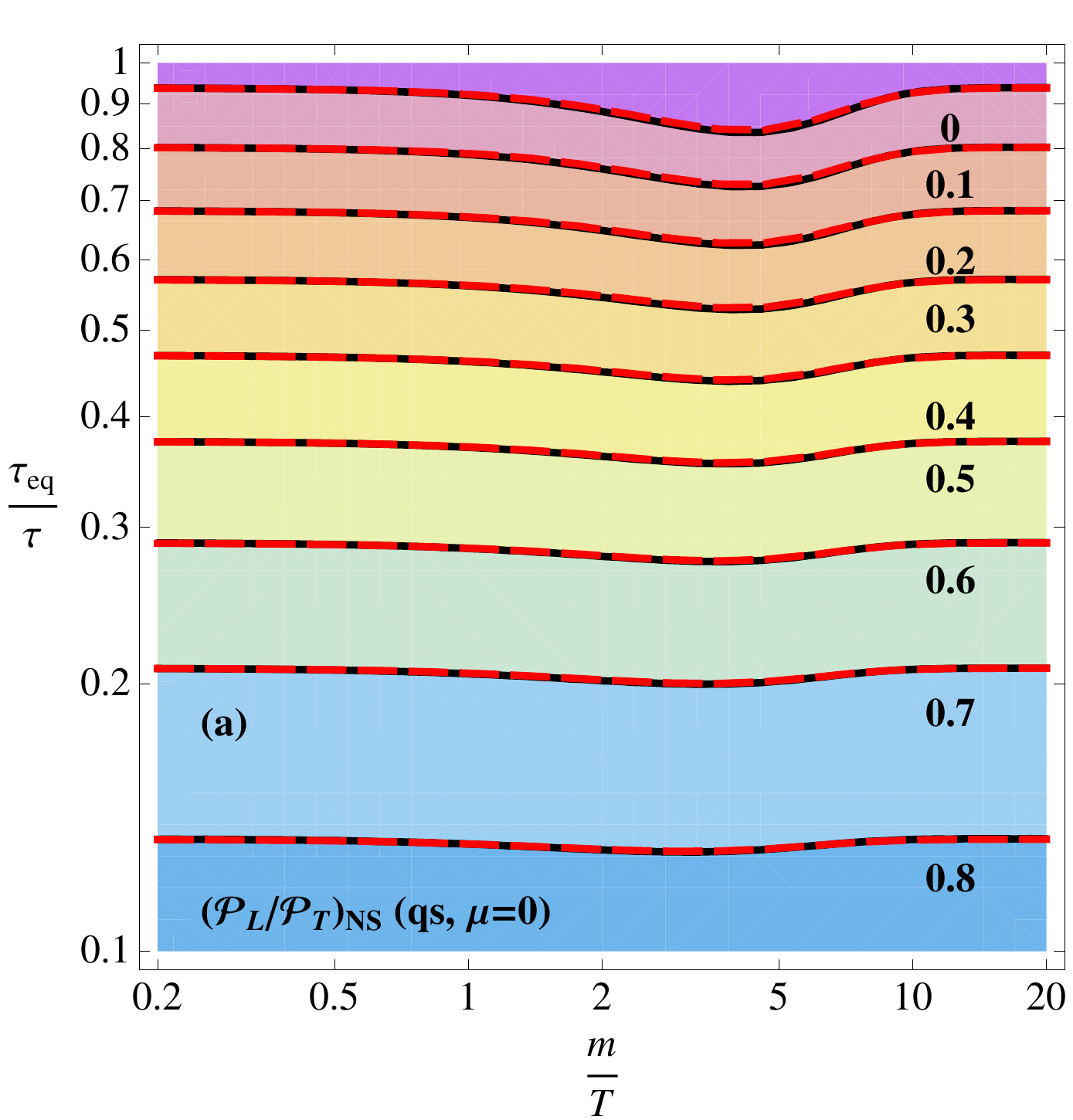}} 
\subfigure{\includegraphics[angle=0,width=0.47\textwidth]{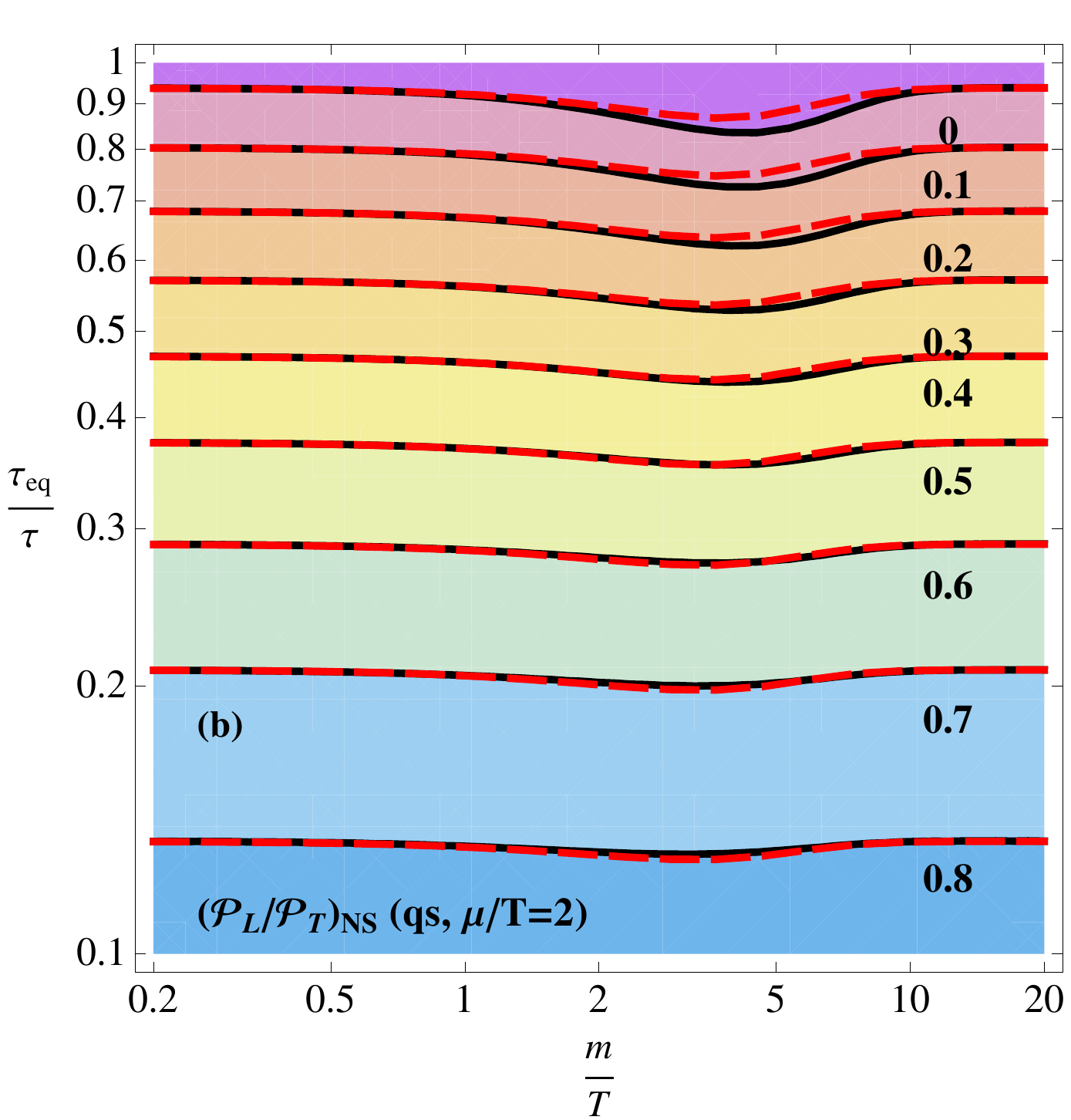}}
\end{center}
\caption{(Color online) Contour plots of $({\cal P}_L/{\cal P}_T)_{\rm NS}$  obtained for the Navier-Stokes hydrodynamics for $\mu = 0$ (a) and $\mu/T=2$ (b). In the two cases quarks and gluons are described by quantum statistics. The solid black lines together with the contour shading represent the classical baryon-free system.}
\label{fig:scaling}
\end{figure}

To analyse the $\left({\cal P}_L/{\cal P}_T\right)_{\rm NS}$ ratio in a general case in Fig.~\ref{fig:scaling} we plot it as a function of  two variables, 
$\teq/\tau$ and $m/T$, for a fixed value of $\mu$.
The left panel of Fig.~\ref{fig:scaling} shows the contour plot of $\left({\cal P}_L/{\cal P}_T\right)_{\rm NS}$ 
in the case where quantum statistics are used and $\mu=0$.
The fact that the contour (red dashed) lines have horizontal shapes indicates that $\left({\cal P}_L/{\cal P}_T\right)_{\rm NS}$ 
depends effectively only on $\teq/\tau$
(except for the region where $\tau \approx \teq$ and $T \approx m/5$). The red dashed lines overlap with solid black lines corresponding to the result for 
the case of classical statistics. It shows that quantum statistics  have negligible effect on $\left({\cal P}_L/{\cal P}_T\right)_{\rm NS}$ 
in the  studied, rather broad range of $\teq/\tau$ and $m/T$. These observations explain similarities of the close-to-equilibrium behavior of $\left({\cal P}_L/{\cal P}_T\right)_{\rm NS}$ in the left panels of Figs. \ref{fig:ooPP}, \ref{fig:opPP}, and \ref{fig:ppPP}. The right panel of 
Fig.~\ref{fig:scaling} shows the contour plot of $\left({\cal P}_L/{\cal P}_T\right)_{\rm NS}$ for $\mu/T=2$. 
In this case we find again a weak dependence on $m/T$ as compared to the case of classical statistics and $\mu=0$ represented by the solid black lines. Again this helps to understand the similarities of the right and left panels of Figs. \ref{fig:ooPP}, \ref{fig:opPP}, and \ref{fig:ppPP}.

\begin{figure}[t]
\begin{center}
\includegraphics[angle=0,width=0.6\textwidth]{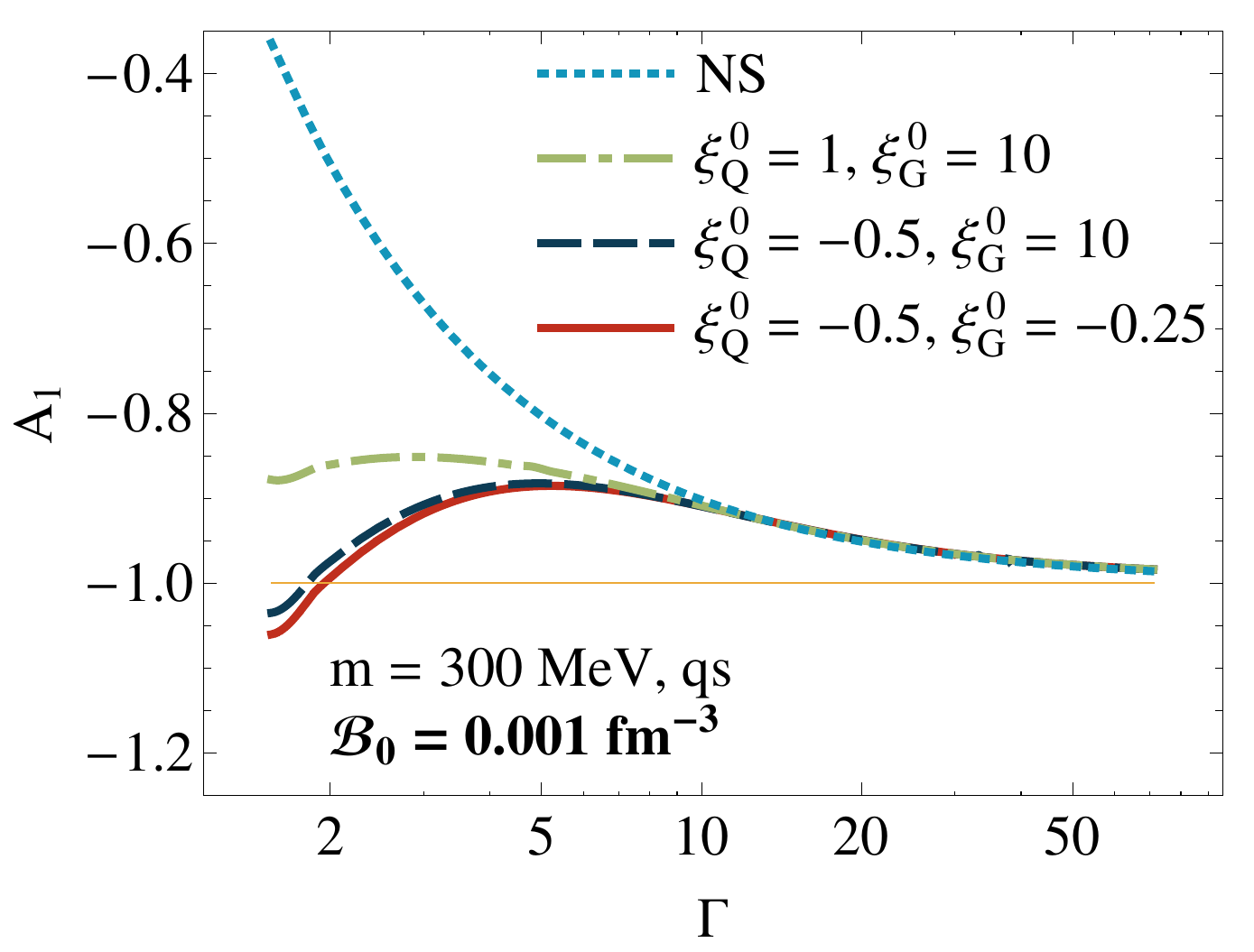}
\end{center}
\caption{(Color online) The quantity $A_1$ plotted as a function of $\Gamma$ for three different initial anisotropies, finite quark mass, and quantum statistics. }
\label{fig:A1}
\end{figure}

\begin{figure}[t]
\begin{center}
\includegraphics[angle=0,width=0.6\textwidth]{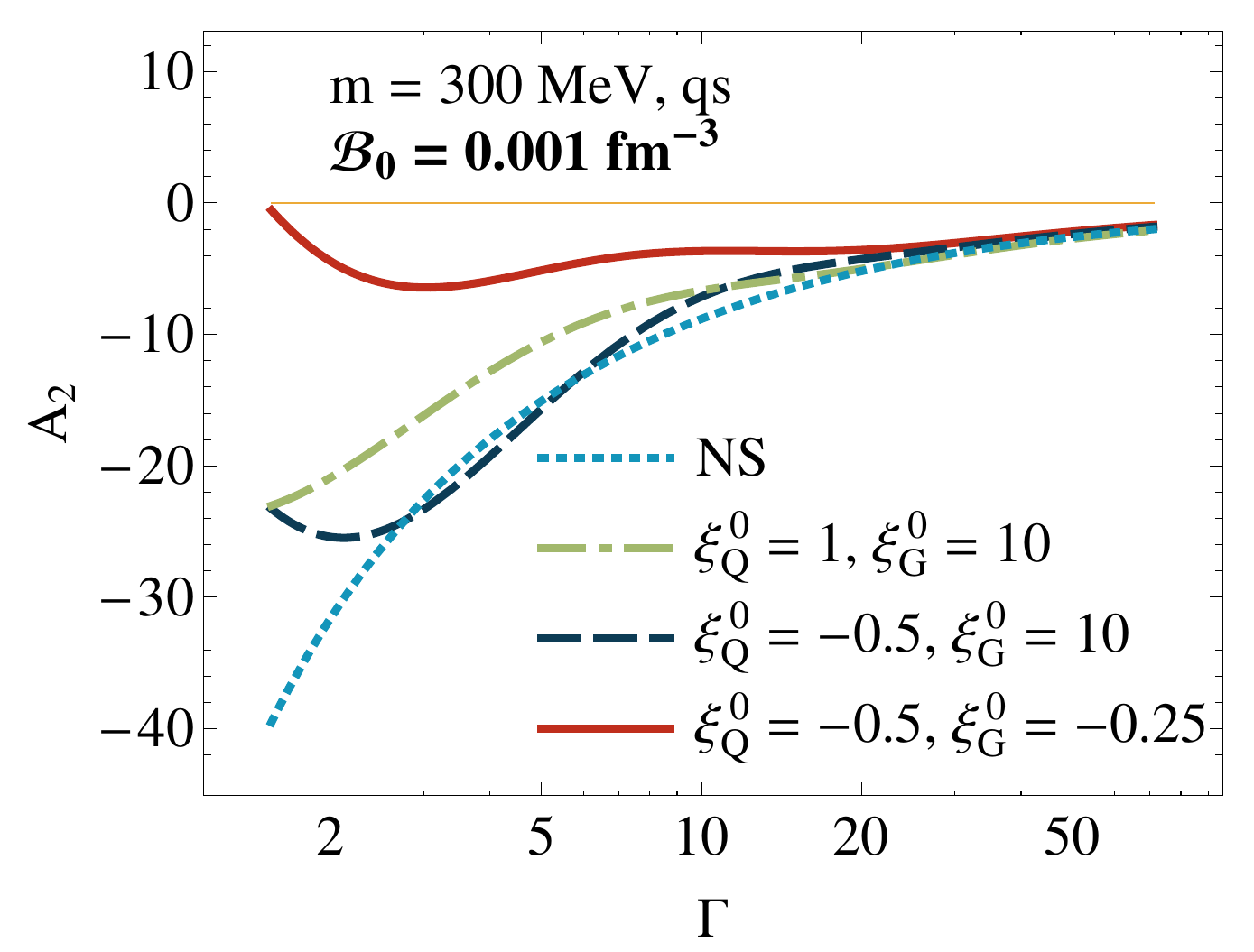}
\end{center}
\caption{(Color online) Same as Fig.~\ref{fig:A1} but for $A_2$ vs. $\Gamma$. }
\label{fig:A2}
\end{figure}

\subsection{Remarks on non-conformal attractor}
\label{sect:nonconfat}

In a very recent paper \cite{Romatschke:2017acs} it has been suggested by Romatschke to look for attractor behaviour by studying the quantities
\bel{A1}
A_1 = \f{\tau d{\cal E}}{({\cal E}+ {\cal P}_{\rm eq})d\tau}
\eel
and
\bel{A2}
A_2 = \f{2 {\cal P}_T + {\cal P}_L - 3 {\cal P}_{\rm eq}}{\zeta T}
\eel
as functions of the variable
\bel{Gamma}
\Gamma = \tau \left[ \f{4}{3} \f{\eta}{{\cal E}+ {\cal P}_{\rm eq}} + \f{\zeta}{{\cal E}+ {\cal P}_{\rm eq}} \right]^{-1}.
\eel
Note that in Eqs. \rfn{A1}--\rfn{Gamma} we used boost invariance to simplify our notation.

In Fig.~\ref{fig:A1} we show the function $A_1(\Gamma)$ obtained for three different initial anisotropies studied in this 
work. To get the connection with \cite{Romatschke:2017acs} we consider the case with negligible baryon number density. Otherwise, we include the 
finite mass of quarks and quantum statistics. Figure ~\ref{fig:A1} shows that the lines corresponding to three different initial
conditions converge and later approach the Navier-Stokes line. This observation supports the existence of a non-conformal attractor
for $A_1$ in our system.

Figure~\ref{fig:A2} shows similar results as Fig.~\ref{fig:A1} but for $A_2(\Gamma)$. In this case, the lines corresponding to
different initial conditions converge with each other only in the NS regime. Hence, our present results are insufficient to 
demonstrate the existence of an attractor for $A_2$.  Further study of this behaviour is planned for our future investigations.
  
\section{Summary and Conclusions}
\label{sect:sumcon}

In this work we have solved a system of coupled kinetic equations for quarks, antiquarks and gluons in the relaxation time
approximation. We have generalised previous results by including: the finite quark mass, the quantum statistics for both
quarks and gluons, and the finite baryon number. We have compared the results of the numerical calculations with the
first-order hydrodynamic calculations to demonstrate the hydrodynamization process. We have found that
equalisation of the longitudinal and transverse pressures takes place earlier than equalisation of the average and equilibrium pressures.
We have determined the shear and bulk viscosities of a mixture and find that the shear viscosity is a sum of the quark and gluon shear viscosities, 
while the bulk viscosity of a mixture is given by the formula known for a massive quark gas. However, the bulk viscosity depends 
on thermodynamic coefficients characterising the whole mixture rather than quarks alone, which means that massless gluon
do contribute to the bulk viscosity (if quarks are massive).

\begin{acknowledgments}
We thank Paul Romatschke, Michal Spalinski, and Michael Strickland for clarifying discussion of our
preliminary results during the Initial Stages 2017 conference in Krakow.
W.F.  and R.R. were supported in part by the Polish National Science Center Grant No. 2016/23/B/ST2/00717.  
\end{acknowledgments}

\appendix
%
\section{Generalized thermodynamic functions}
\label{s:tf}

In this section we present explicit expressions for various physical quantities such as the particle and energy densities or the transverse  and longitudinal pressures. These expressions are obtained with the use of different distribution functions
which not necessarily correspond to local equilibrium. Thus, we call them generalised thermodynamic functions --- in local equilibrium they become standard thermodynamic functions satisfying well known thermodynamic identities.  We start with the anisotropic RS distributions, as other cases can be easily worked out if the results for the RS distributions are known.

\subsection{Anisotropic distributions}
\label{ss:ae}
%
The forms of the generalised thermodynamic functions  for anisotropic distributions are given by the following integrals:
\begin{eqnarray}
{\cal N}^{\s,\an}  \equiv n_{U}^{\s, \an} &=&  k_\s   \int \!dP \, \lp p\cdot U \rp  f_{\s, \an}\lsb\VP\,  p \cdot U, p \cdot Z \rsb   ,
\label{eq:nans}\\
{\cal E}^{\s,\an} \equiv t_{UU}^{\s, \an} &=& k_\s\int dP \, \lp p\cdot U \rp^2 f_{\s, \an}\lsb\VP\,  p \cdot U, p \cdot Z\rsb ,
\label{eq:eans}\\
{\cal P}^{\s,\an}_T \equiv t_{AA}^{\s, \an} 
&=& k_\s\int  dP \, \lp p\cdot X \rp^2 f_{\s, \an}\lsb\VP\,  p \cdot U, p \cdot Z\rsb  \qquad (A\neq U, Z) \nn\\
&=& k_\s\int  dP \, \lp p\cdot Y \rp^2 f_{\s, \an}\lsb\VP\,  p \cdot U, p \cdot Z\rsb \nn\\
&=& - \frac{k_\s}{2}\int dP \, (p\cdot \Delta_T \cdot p)f_{\s, \an}\lsb\VP\,  p \cdot U, p \cdot Z\rsb ,
\label{eq:ptans}\\
{\cal P}^{\s,\an}_L \equiv t_{ZZ}^{\s, \an} &=& k_\s\int  dP \, \lp p\cdot Z \rp^2 f_{\s, \an}\lsb\VP\,  p \cdot U, p \cdot Z\rsb  .
\label{eq:plans}
\end{eqnarray}
The explicit calculations lead to the following expressions  for quarks and antiquarks
\begin{eqnarray}
{\cal N}^{\Qpm, \an}    &=&   4 \pi k_\Q  \Lambda_\Q^3 \tilde{{\cal H}}_{\cal N}^+\lp \f{1}{\sqrt{1+\xi_\Q}}, \frac{m}{\Lambda_\Q}, \mp \frac{\lambda}{\Lambda_\Q}\rp,
\label{eq:nan}\\
 {\cal E}^{\Qpm, \an}   &=&    2 \pi k_\Q    \Lambda_\Q^4 \tilde{{\cal H}}^+\lp \f{1}{\sqrt{1+\xi_\Q}}, \frac{m}{\Lambda_\Q}, \mp \frac{\lambda}{\Lambda_\Q}\rp,
\label{eq:ean}\\
{\cal P}^{\Qpm,\an}_T   &=&     \pi k_\Q   \Lambda_\Q^4 \tilde{{\cal H}}_T^+\lp \f{1}{\sqrt{1+\xi_\Q}}, \frac{m}{\Lambda_\Q}, \mp \frac{\lambda}{\Lambda_\Q}\rp,
\label{eq:ptan}\\
{\cal P}^{\Qpm,\an}_L   &=&   2 \pi k_\Q   \Lambda_\Q^4 \tilde{{\cal H}}_L^+\lp \f{1}{\sqrt{1+\xi_\Q}}, \frac{m}{\Lambda_\Q}, \mp \frac{\lambda}{\Lambda_\Q}\rp,
\label{eq:plan}
\end{eqnarray}
 where functions $\tilde{{\cal H}}$ are defined by the integrals:
\beal{Ht}
\tilde{{\cal H}}_{\cal N}^\pm\lp a,y,z\rp &\equiv& \int\limits_0^\infty r^2  dr \, h^{^\pm}_\eq\lp \sqrt{r^2+y^2}+z\rp a,  \\
\tilde{{\cal H}} ^\pm\lp a,y,z\rp &\equiv& \int\limits_0^\infty r^3  dr \,  h^{^\pm}_\eq\lp \sqrt{r^2+y^2}+z\rp   {\cal H}_{2}\lp a,\frac{y}{r} \rp, \nn\\
\tilde{{\cal H}}_T^\pm\lp a,y,z\rp &\equiv& \int\limits_0^\infty r^3  dr \,  h^{^\pm}_\eq\lp \sqrt{r^2+y^2}+z\rp   {\cal H}_{2T}\lp a,\frac{y}{r} \rp, \nn\\
\tilde{{\cal H}}_L ^\pm\lp a,y,z\rp &\equiv& \int\limits_0^\infty r^3  dr  \, h^{^\pm}_\eq\lp \sqrt{r^2+y^2}+z\rp   {\cal H}_{2L}\lp a,\frac{y}{r} \rp, \nn
\eeal
and the functions ${\cal H}_2(a,b)$ were introduced in~\cite{Florkowski:2014sfa}:
\beal{H2}
{\cal H}_{2}\lp a,b \rp &\equiv& a  \int \limits_0^\pi   d\varphi\,\sin\varphi\sqrt{ a^2 \cos^2\varphi +\sin^2\varphi +b^2},\\ 
{\cal H}_{2T}\lp a,b \rp &\equiv& a \int\limits_0^{\pi} d\varphi\,
\f{\sin^3\varphi  }{\sqrt{a^2\cos^2\varphi +\sin^2\varphi +b^2}},\nn\\ 
{\cal H}_{2L}\lp a,b \rp &\equiv& a^3\int\limits_0^{\pi} d\varphi\,
\f{\sin\varphi \,\cos^2\varphi}{\sqrt{a^2\cos^2\varphi +\sin^2\varphi +b^2}}.\nn 
\eeal
With $b=0$ the functions ${\cal H}_2(a,b)$ reduce to the functions ${\cal H}(a), {\cal H}_{L}(a)$ and ${\cal H}_{T}(a)$ used in~\cite{Florkowski:2013lya}. The integrals in \EQB{H2} are analytic~\cite{Florkowski:2014sfa}:
\bea\label{eq:H2}
\hspace{-1cm} {\cal H}_2(a,b)  &=& \frac{a}{\sqrt{a^2-1}} \left( (1+b^2)
\tanh^{-1} \sqrt{\frac{a^2-1}{a^2+b^2}} + \sqrt{(a^2-1)(a^2+b^2)} \, \right),\\
\hspace{-1cm} {\cal H}_{2T}(a,b)  
&=& \frac{a}{(a^2-1)^{3/2}}
\lp\left(b^2+2a^2-1\right) 
\tanh^{-1}\sqrt{\frac{a^2-1}{a^2+b^2}}
-\sqrt{(a^2-1)(a^2+b^2)} \rp,  
\label{eq:H2T}\\
\hspace{-1cm} {\cal H}_{2L}(a,b)  
&=& \frac{a^3}{(a^2-1)^{3/2}}
\lp-(1+b^2)
\tanh^{-1}\sqrt{\frac{a^2-1}{a^2+b^2}}
+\sqrt{(a^2-1)(a^2+b^2)} \,\,\rp. 
\label{eq:H2L}
\eea
For gluons one has:  
\begin{eqnarray}
{\cal N}^{\G, \an}  &=&  4 \pi k_\G  \Lambda_\G^3 \tilde{{\cal H}}_{\cal N}^-\lp \f{1}{\sqrt{1+\xi_\G}}, 0,0\rp = 8 \pi \zeta(3)  k_\G  \frac{\Lambda_\G^3}{\sqrt{1+\xi_\G}} ,
\label{eq:nang}\\
 {\cal E}^{\G, \an} &=&  2 \pi k_\G \Lambda_\G^4 \tilde{{\cal H}}^-\lp \f{1}{\sqrt{1+\xi_\G}}, 0,0\rp =\frac{2  \pi^5}{15} k_\G \Lambda_\G^4  {\cal H}  \lp \f{1}{\sqrt{1+\xi_\G}} \rp,
\label{eq:eang}\\
{\cal P}^{\G,\an}_T  &=&   \pi k_\G   \Lambda_\G^4 \tilde{{\cal H}}_T^-\lp \f{1}{\sqrt{1+\xi_\G}}, 0,0\rp =\frac{ \pi^5}{15} k_\G \Lambda_\G^4  {\cal H}_T  \lp \f{1}{\sqrt{1+\xi_\G}} \rp,
\label{eq:ptang}\\
{\cal P}^{\G,\an}_L  &=& 2 \pi k_\G   \Lambda_\G^4 \tilde{{\cal H}}_L^-\lp \f{1}{\sqrt{1+\xi_\G}}, 0,0\rp =\frac{ 2 \pi^5}{15} k_\G \Lambda_\G^4  {\cal H}_L  \lp \f{1}{\sqrt{1+\xi_\G}} \rp,
\label{eq:plang}
\end{eqnarray}
where $\zeta$ is the Riemann zeta function (the coefficient $\zeta(3)$ is known as Ap\'ery's constant). The expressions on the right-hand sides of
Eqs.~\rfn{eq:nang}--\rfn{eq:plang} hold for the Bose-Einstein statistics. Note that in the case of massless gluons the integrals \rfn{H2} are done for $b=0$ and can be factorized in~\EQS{Ht}. 

It is useful to notice that the functions ${\cal H}_{2}$ and ${\cal H}_{2L}$ are related by the expression
\bel{dH2}
\f{\partial {\cal H}_{2}\lp a,b \rp}{\partial a} = \f{{\cal H}_{2}\lp a,b \rp + {\cal H}_{2L}\lp a,b \rp}{a},
\eel
hence, we also have
\bel{dtH2}
\f{\partial \tilde{{\cal H}} ^\pm\lp a,y,z\rp }{\partial a} = \f{\tilde{{\cal H}} ^\pm\lp a,y,z\rp + \tilde{{\cal H}} ^\pm\lp a,y,z\rp}{a}.
\eel
We can use \rfn{dtH2} to derive \rfn{SECOND-EQUATION-1} from \rfn{SECOND-EQUATION}.

We close this section with the formula for the baryon number density valid for anisotropic RS systems
\begin{eqnarray}
{\cal B}^\an &=& \frac{{\cal N}^{\Qp, \an} -{\cal N}^{\Qm, \an}}{3} 
=   \frac{16 \pi k_\Q \Lambda_\Q^3}{3 \sqrt{1+\xi_\Q}}  \sinh\lp\frac{\lambda}{\Lambda_\Q}\rp\,{\cal H}_{\cal B}\lp \frac{m}{\Lambda_\Q},   \frac{\lambda}{\Lambda_\Q}\rp,
\end{eqnarray} 
where
\begin{eqnarray}
{\cal H}_{\cal B}\lp y,z\rp &\equiv&
  \frac{1}{4}\int\limits_0^\infty r^2  dr \lsb  \frac{1}{ \cosh \sqrt{r^2+y^2}+\cosh z   } \rsb .
\label{HB}
\end{eqnarray}
%

\subsection{Isotropic distributions}
\label{ss:ie}
%
The forms of the thermodynamic functions for the equilibrium state are commonly known, nevertheless, we quote them here for completeness. They are given by the formulas
\begin{eqnarray}
{\cal N}^{\s, \eq} \equiv n_{U}^{\s, \eq} &=&  k_\s   \int \! dP \, \lp p\cdot U \rp f_{\s, \eq}(p\cdot U)   ,
\label{eq:neq}\\
 {\cal E}^{\s, \eq}\equiv t_{UU}^{\s, \eq} &=& k_\s\int dP \, \lp p\cdot U \rp^2 f_{\s, \eq}(p\cdot U), 
\label{eq:eeq}\\
{\cal P}^{\s, \eq}\equiv t_{AA}^{\s, \eq} &=& k_\s\int \!dP\, \lp p\cdot A \rp^2 f_{\s, \eq}(p\cdot U)  \nn\\
&=& -\frac{k_\s}{3}\int \!dP\, (p\cdot \Delta \cdot p) f_{\s, \eq}(p\cdot U),  \quad (A\neq U) .
\label{eq:peq}
\end{eqnarray}
Their explicit forms for quarks and anti-quarks may be obtained  from  \EQSM{eq:nan}{eq:plan} as a special case of $\xi_\s\to 0$, $\Lambda_\s\to T$, and $\lambda_\s\to \mu$,
\begin{eqnarray}
{\cal N}^{\Qpm, \eq} \!\!  &=&\!\!  4 \pi k_\Q  T^3 \tilde{{\cal H}}_{\cal N}^+\lp 1, \frac{m}{T}, \mp \frac{\mu}{T}\rp,
\label{eq:nqeq}\\
 {\cal E}^{\Qpm, \eq} \!\! &=&\!\!   2 \pi k_\Q    T^4 \tilde{{\cal H}}^+\lp 1, \frac{m}{T}, \mp \frac{\mu}{T}\rp,
\label{eq:eqeq}\\
{\cal P}^{\Qpm,\eq}_T  \!\! &=&\!\!    \pi k_\Q   T^4 \tilde{{\cal H}}_T^+\lp 1, \frac{m}{T}, \mp \frac{\mu}{T}\rp,
\label{eq:ptqeq}\\
{\cal P}^{\Qpm,\eq}_L  \!\! &=&\!\!  2 \pi k_\Q   T^4 \tilde{{\cal H}}_L^+\lp 1, \frac{m}{T}, \mp \frac{\mu}{T}\rp.
\label{eq:plqeq}
\end{eqnarray}
Note that ${\cal H}_{2L}(1,b)=2/\lp 3\sqrt{1+b^2} \rp$ and ${\cal H}_{2T}(1,b)=4/\lp 3\sqrt{1+b^2} \rp$ which means that ${\cal P}^{\Qpm,\eq}_T ={\cal P}^{\Qpm,\eq}_L \equiv {\cal P}^{\Qpm,\eq}$, as expected for the isotropic state.  
%
Analogous results may be obtained for gluons  
\begin{eqnarray}
{\cal N}^{\G, \eq}  \!\!&=&\!\!  4 \pi k_\G  T^3 \tilde{{\cal H}}_{\cal N}^-\lp 1, 0,0\rp = 8  \pi \zeta(3) k_\G  T^3 ,
\label{eq:ngeq}\\
 {\cal E}^{\G, \eq} \!\!&=&\!\!  2 \pi k_\G T^4 \tilde{{\cal H}}^-\lp 1, 0,0\rp=\frac{4  \pi^5}{15} k_\G T^4 ,
\label{eq:egeq}\\
{\cal P}^{\G,\eq}_T  \!\!&=&\!\!   \pi k_\G   T^4 \tilde{{\cal H}}_T^-\lp 1, 0,0\rp=\frac{4  \pi^5}{45} k_\G T^4,
\label{eq:ptgeq}\\
{\cal P}^{\G,\eq}_L  \!\!&=&\!\! 2 \pi k_\G   T^4 \tilde{{\cal H}}_L^-\lp 1, 0,0\rp=\frac{4  \pi^5}{45} k_\G T^4,
\label{eq:plgeq}
\end{eqnarray}
where to get the last expressions on the right-hand sides we again assumed the Bose-Einstein statistics. Here, similarly as for quarks ${\cal P}^{\G,\eq}_T ={\cal P}^{\G,\eq}_L \equiv {\cal P}^{\G,\eq}$. 
Similarly to anisotropic case the baryon number density is
\begin{eqnarray}
{\cal B}^\eq &=& \frac{{\cal N}^{\Qp, \eq} -{\cal N}^{\Qm, \eq}}{3} 
= \frac{16 \pi k_\Q T^3}{3} \sinh\lp\frac{\mu}{T}\rp\,{\cal H}_{\cal B}\lp \frac{m}{T},   \frac{\mu}{T}\rp.
\label{BeqApp}
\end{eqnarray} 
%
%
\subsection{Exact solution of the kinetic equations}
\label{ss:esotke}
%
For the solutions of \EQS{BIke} of the form \EQB{formsolQ} the thermodynamic variables have the forms
\begin{eqnarray}
{\cal N}^{\s}  &=&  k_\s   \int \!dP \, \lp p\cdot U \rp  f_{\s}\lsb\VP\,  p \cdot U, p \cdot Z \rsb   \nn,
\label{eq:nane}\\
{\cal E}^{\s}  &=& k_\s\int dP \, \lp p\cdot U \rp^2 f_{\s}\lsb\VP\,  p \cdot U, p \cdot Z\rsb ,
\label{eq:eane}\\
{\cal P}^{\s}_T  &=& k_\s\int  dP \, \lp p\cdot A \rp^2 f_{\s}\lsb\VP\,  p \cdot U, p \cdot Z\rsb  \qquad  (A\neq U,Z), \nn
\label{eq:ptane}\\
{\cal P}^{\s}_L  &=& k_\s\int  dP \, \lp p\cdot Z \rp^2 f_{\s}\lsb\VP\,  p \cdot U, p \cdot Z\rsb. \nn
\label{eq:plane}
\end{eqnarray}
Using above definitions and repeating the calculation from \SEC{ss:ae}, gives 
\begin{eqnarray}
{\cal N}^{\Qpm}  &=&  4 \pi k_\Q \lsb  \lp\Lambda_\Q^0\rp^3 \tilde{{\cal H}}_{\cal N}^+\lp \f{\tau_0}{\tau \sqrt{1+\xi_\Q^0}}, \frac{m}{\Lambda_\Q^0}, \mp \frac{\lambda^0}{\Lambda_\Q^0}\rp D(\tau,\tau_0) \right. \label{eq:nq} \\
&&  \left. \hspace{5cm} + \int\l_{\tau_0}^{\tau} \f{d \tau'}{\teq^\prime}\  D(\tau,\tau') \lp T^\prime\rp^3 \tilde{{\cal H}}_{\cal N}^+\lp \f{\tau^\prime}{\tau  }, \frac{m}{T^\prime}, \mp \frac{\mu^\prime}{T^\prime}\rp\rsb, \nn
\end{eqnarray}
\begin{eqnarray}
 {\cal E}^{\Qpm} &=&  2 \pi k_\Q \lsb  \lp\Lambda_\Q^0\rp^4 \tilde{{\cal H}}^+\lp \f{\tau_0}{\tau \sqrt{1+\xi_\Q^0}}, \frac{m}{\Lambda_\Q^0}, \mp \frac{\lambda^0}{\Lambda_\Q^0}\rp D(\tau,\tau_0) \right. \label{eq:eq} \\
&&  \left. \hspace{5cm} + \int\l_{\tau_0}^{\tau} \f{d \tau'}{\teq^\prime}\  D(\tau,\tau') \lp T^\prime\rp^4 \tilde{{\cal H}}_{\cal  }^+\lp \f{\tau^\prime}{\tau  }, \frac{m}{T^\prime}, \mp \frac{\mu^\prime}{T^\prime}\rp\rsb, \nn
\end{eqnarray}
\begin{eqnarray}
{\cal P}^{\Qpm}_T  &=&   \pi k_\Q  \lsb \lp\Lambda_\Q^0\rp^4 \tilde{{\cal H}}_T^+\lp \f{\tau_0}{\tau \sqrt{1+\xi_\Q^0}}, \frac{m}{\Lambda_\Q^0}, \mp \frac{\lambda^0}{\Lambda_\Q^0}\rp D(\tau,\tau_0) \right. \label{eq:ptq} \\
&&  \left. \hspace{5cm} + \int\l_{\tau_0}^{\tau} \f{d \tau'}{\teq^\prime}\  D(\tau,\tau')\lp T^\prime\rp^4 \tilde{{\cal H}}_{  T}^+\lp \f{\tau^\prime}{\tau  }, \frac{m}{T^\prime}, \mp \frac{\mu^\prime}{T^\prime}\rp \rsb, \nn
\end{eqnarray}
\begin{eqnarray}
{\cal P}^{\Qpm}_L  &=& 2 \pi k_\Q  \lsb \lp\Lambda_\Q^0\rp^4 \tilde{{\cal H}}_L^+\lp \f{\tau_0}{\tau \sqrt{1+\xi_\Q^0}}, \frac{m}{\Lambda_\Q^0}, \mp \frac{\lambda^0}{\Lambda_\Q^0}\rp D(\tau,\tau_0) \right. \label{eq:plq} \\
&&  \left. \hspace{5cm} + \int\l_{\tau_0}^{\tau} \f{d \tau'}{\teq^\prime}\  D(\tau,\tau') \lp T^\prime\rp^4 \tilde{{\cal H}}_{L}^+\lp \f{\tau^\prime}{\tau  }, \frac{m}{T^\prime}, \mp \frac{\mu^\prime}{T^\prime}\rp\rsb, \nn
\end{eqnarray}
for quarks and 
\begin{eqnarray}
{\cal N}^{\G}  &=&  4 \pi k_\G \lsb  \lp\Lambda_\G^0\rp^3 \tilde{{\cal H}}_{\cal N}^-\lp \f{\tau_0}{\tau \sqrt{1+\xi_\G^0}}, 0,0\rp D(\tau,\tau_0)+ \int\l_{\tau_0}^{\tau} \f{d \tau'}{\teq^\prime}\  D(\tau,\tau') \lp T^\prime\rp^3 \tilde{{\cal H}}_{\cal N}^-\lp \f{\tau^\prime}{\tau  }, 0,0\rp\rsb, 
\nn \\
\label{eq:ng}
\end{eqnarray}
\begin{eqnarray}
 {\cal E}^{\G} &=&  2 \pi k_\G \lsb  \lp\Lambda_\G^0\rp^4 \tilde{{\cal H}}^-\lp \f{\tau_0}{\tau \sqrt{1+\xi_\G^0}},  0,0\rp D(\tau,\tau_0)+ \int\l_{\tau_0}^{\tau} \f{d \tau'}{\teq^\prime}\  D(\tau,\tau') \lp T^\prime\rp^4 \tilde{{\cal H}}_{\cal  }^-\lp \f{\tau^\prime}{\tau  }, 0,0\rp\rsb,
 \nn \\
\label{eq:eg}
\end{eqnarray}
\begin{eqnarray}
{\cal P}^{\G}_T  &=&   \pi k_\G  \lsb \lp\Lambda_\G^0\rp^4 \tilde{{\cal H}}_T^-\lp \f{\tau_0}{\tau \sqrt{1+\xi_\G^0}},  0,0\rp D(\tau,\tau_0)+ \int\l_{\tau_0}^{\tau} \f{d \tau'}{\teq^\prime}\  D(\tau,\tau')\lp T^\prime\rp^4 \tilde{{\cal H}}_{  T}^-\lp \f{\tau^\prime}{\tau  }, 0,0\rp \rsb,
\nn \\
\label{eq:ptg}
\end{eqnarray}
\begin{eqnarray}
{\cal P}^{\G}_L  &=& 2 \pi k_\G  \lsb \lp\Lambda_\G^0\rp^4 \tilde{{\cal H}}_L^-\lp \f{\tau_0}{\tau \sqrt{1+\xi_\G^0}},  0,0\rp D(\tau,\tau_0)+ \int\l_{\tau_0}^{\tau} \f{d \tau'}{\teq^\prime}\  D(\tau,\tau') \lp T^\prime\rp^4 \tilde{{\cal H}}_{L}^-\lp \f{\tau^\prime}{\tau  }, 0,0\rp\rsb,
\nn \\
\label{eq:plg}
\end{eqnarray}
for gluons.
We define the baryon number density for the exact solution of the kinetic equation as follows
\begin{eqnarray}
{\cal B}  &=&  \frac{16 \pi k_\Q}{3} \lsb   \f{\tau_0 \lp\Lambda_\Q^0\rp^3}{\tau \sqrt{1+\xi_\Q^0}} \sinh\lp\frac{\lambda^0}{\Lambda_\Q^0}\rp\,{\cal H}_{\cal B}\lp \frac{m}{\Lambda_\Q^0},\frac{\lambda^0}{\Lambda_\Q^0} \rp D(\tau,\tau_0) \right. \label{BApp} \\
&& \left. \hspace{5cm}
+ \int\l_{\tau_0}^{\tau} \f{d \tau'}{\teq^\prime}\  D(\tau,\tau')  \f{\tau^\prime \lp T^\prime\rp^3}{\tau  } \sinh\lp\frac{\mu^\prime}{T^\prime}\rp\,{\cal H}_{\cal B}\lp \frac{m}{T^\prime}, \frac{\mu^\prime}{T^\prime}\rp\rsb  . \nn 
\end{eqnarray} 

%
\section{Navier-Stokes hydrodynamics}
\label{s:NS}
%

The results of our kinetic-theory calculations are compared with the viscous hydrodynamic results obtained by solving 
the Navier-Stokes (NS) hydrodynamic equations. The latter have the form
\begin{eqnarray}
\f{d}{d\tau} \left( {\cal E}^{\Q, \eq}  +  {\cal E}^{\G, \eq}  \right)
&=& - \f{{\cal E}^{\Q, \eq} +  {\cal E}^{\G, \eq} + {\cal P}^{\Q,\eq}  +  {\cal P}^{\G,\eq}  + \Pi_{\rm NS} - \pi_{\rm NS}  }{\tau}, 
\label{NSapp1} \\
\f{d{\cal B}^{\eq}}{d\tau} +\f{{\cal B}^{\eq}}{\tau} &=& 0.
\label{NSapp2}
\end{eqnarray}
Here ${\cal E}^{\Q, \eq}  = {\cal E}^{\Qp, \eq} +  {\cal E}^{\Qm, \eq}$ is the equilibrium energy density of quarks and antiquarks, ${\cal P}^{\Q,\eq} = {\cal P}^{\Qp,\eq} + {\cal P}^{\Qm,\eq}$ is the equilibrium pressure of quarks and antiquarks, $\Pi_{\rm NS}$ is the bulk pressure, and $\pi_{\rm NS}$ is the shear pressure (both used in the close-to-equilibrium limit). All the functions appearing in \rfn{NSapp1} and \rfn{NSapp2} depend on $T$ and $\mu$, hence  Eqs.~\rfn{NSapp1} and \rfn{NSapp2} are two coupled equations that can be used to determine $T(\tau)$ and $\mu(\tau)$.  
One can easily notice that Eq.~\rfn{NSapp1} may be written in the form of Eq.~\rfn{SECOND-EQUATION-1} once we identify
\beal{PLPTns}
 {\cal P}_L   &=&  {\cal P}^{\rm eq} - \pi_{\rm NS} + \Pi_{\rm NS},\\
 {\cal P}_T  &=&  {\cal P}^{\rm eq} + \frac{1}{2}\pi_{\rm NS} + \Pi_{\rm NS}, \nn
\eeal
where ${\cal P}^{  \eq}  = {\cal P}^{\Q, \eq} +  {\cal P}^{\G, \eq}$.
Within NS approach, the shear and bulk pressures  are expressed by the kinetic coefficients $\eta$ and $\zeta$,
\bel{pi2}
\pi_{\rm NS} = \f{4 \eta}{3\tau} = \f{4 (\eta_\Q+\eta_G)}{3\tau},
\eel
\bel{Zeta1}
\Pi_{\rm NS} = -\f{\zeta}{\tau}.
\eel
The expressions for $\eta_Q$, $\eta_G$, and $\zeta$ are given by Eqs.~\rfn{etaQ}, \rfn{etaG}, and \rfn{zeta}.

\medskip
For the moment let us denote $T(\tau)$ and $\mu(\tau)$ obtained from the kinetic theory as $T_{\rm KT}(\tau)$  and $\mu_{\rm KT}(\tau)$, while those obtained from the NS hydrodynamics as $T_{\rm NS}(\tau)$ and $\mu_{\rm NS}(\tau)$. We expect that $T_{\rm KT}(\tau)$ and  $\mu_{\rm KT}(\tau)$ agree well with $T_{\rm NS}(\tau)$ and $\mu_{\rm NS}(\tau)$ in the late stages of the evolution, when the system approaches local equilibrium. To check this behaviour we choose such initial conditions for hydrodynamic equations  \rfn{NSapp1} and \rfn{NSapp2} that for the final time $\tau=\tau_f$ we match the temperature and chemical potential in the two approaches: $T_{\rm NS}(\tau_f)=T_{\rm KT}(\tau_f)$,  $\mu_{\rm NS}(\tau_f)=\mu_{\rm KT}(\tau_f)$. Then, we check if the functions  $T_{\rm NS}(\tau)$ and $\mu_{\rm NS}(\tau)$
smoothly approach $T_{\rm KT}(\tau)$  and $\mu_{\rm KT}(\tau)$ if $\tau \to \tau_f$. By neglecting the bulk and shear pressures in   \rfn{NSapp1}  we can also make comparison with perfect fluid hydrodynamics and check if the system approaches local equilibrium.

\section{Shear and Bulk viscosities for mixtures}
\label{s:shearbulk}

In this section we present details of our method used to calculate the shear and bulk viscosity coefficients for a quark-gluon mixture. 
We follow the treatment of Refs.~\cite{Florkowski:2015rua,Florkowski:2015dmm}, where the bulk viscosity was obtained 
for the Gribov-Zwanziger plasma.  Analyzing a boost-invariant system, we deal with a simple structure of hydrodynamic equations, 
which facilitates the calculations. 

\subsection{Landau matching conditions in the case of boost invariant geometry}
\label{s:LM-BI}

In the first-order gradient expansion, the non-equilibrium corrections to the equilibrium distribution function
have the form
\bel{df}
\delta f_\Qpm = -\teq \f{\partial f_{\Qpm, \eq}}{\partial \tau}, \quad \delta f_\G = -\teq \f{\partial f_{\G, \eq}}{\partial \tau}.
\eel
Using the form of $f_{\Qpm, \eq}$ and $f_{\G, \eq}$ for boost-invariant geometry we find
\begin{eqnarray}
\delta f_\Qpm &=& -\teq f_{\Qpm, \eq} \left(1 - f_{\Qpm, \eq} \right)
 \left[ \f{w^2}{v \tau^2 T} \pm \f{d\mu}{T d\tau } + \left( \f{v}{\tau} \mp \mu \right) \f{d \ln T}{T d\tau} \right] ,
 \label{df1q} \\
 \delta f_\G &=& -\teq f_{\G, \eq} \left(1 + f_{\G, \eq} \right)
 \left[ \f{w^2}{v \tau^2 T}  + \f{v}{\tau}  \f{d \ln T}{T d\tau}  \right]. \label{df1g}
\end{eqnarray}
The Landau matching conditions for the energy and momentum read
\begin{eqnarray}
&&\int  \f{dw d^2 \pT}{v} \,\f{v^2}{\tau^2} \left[ k_\Q \left( \delta f_\Qp + \delta f_\Qm \right) + k_\G  \delta f_\G \right] = 0 \label{LMdq}, \\
&& \int \f{dw d^2 \pT}{v} \,\f{v^2}{3 \tau^2} \left[ k_\Q \left( \delta f_\Qp - \delta f_\Qm \right)  \right] = 0 \label{LMdg}.
\end{eqnarray}
Using \rfn{df1q} and \rfn{df1g} we rewrite \rfn{LMdq} and \rfn{LMdg} as
\begin{eqnarray}
&&  \int \f{dw d^2 \pT}{v} \, \f{v^2}{\tau^2} \, f_{3S} \, \f{w^2}{v \tau^2 T}
+  \int \f{dw d^2 \pT}{v} \, \f{v^2}{\tau^2} \, f_{2D} \, \f{d\mu}{T d\tau} \nn \\
&& +   \int \f{dw d^2 \pT}{v} \, \f{v^2}{\tau^2} \, f_{3S} \, \f{v}{T \tau} \, \f{d \ln T}{d\tau}
-  \int \f{dw d^2 \pT}{v} \, \f{v^2}{\tau^2} \, f_{2D} \, \f{\mu}{T} \f{d \ln T}{d\tau} = 0
\label{LMdq1}
\end{eqnarray}
and
\begin{eqnarray}
&&  \int \f{dw d^2 \pT}{v} \, \f{v^2}{\tau^2} \, f_{2D} \, \f{w^2}{v \tau^2 T}
+  \int \f{dw d^2 \pT}{v} \, \f{v^2}{\tau^2} \, f_{2S} \, \f{d\mu}{T d\tau} \nn \\
&& +   \int \f{dw d^2 \pT}{v} \, \f{v^2}{\tau^2} \, f_{2D} \, \f{v}{T \tau} \, \f{d \ln T}{d\tau}
-  \int \f{dw d^2 \pT}{v} \, \f{v^2}{\tau^2} \, f_{2S} \, \f{\mu}{T} \f{d \ln T}{d\tau} = 0,
\label{LMdg1}
\end{eqnarray}
where
\begin{eqnarray}
f_{3S} &=& k_\Q \left[ f_{\Qp, \eq} \left(1 - f_{\Qp, \eq} \right) + f_{\Qm, \eq} \left(1 - f_{\Qm, \eq} \right) \right] + k_\G f_{\G, \eq} \left(1 + f_{\G, \eq} \right), \nn \\
f_{2S} &=& k_\Q \left[ f_{\Qp, \eq} \left(1 - f_{\Qp, \eq} \right) + f_{\Qm, \eq} \left(1 - f_{\Qm, \eq} \right) \right] , \nn \\
f_{2D} &=& k_\Q \left[ f_{\Qp, \eq} \left(1 - f_{\Qp, \eq} \right) - f_{\Qm, \eq} \left(1 - f_{\Qm, \eq} \right) \right].
\end{eqnarray}
By introducing the ``averaged'' values defined as
\bel{av}
\langle ... \rangle_\alpha \equiv \int \f{dw d^2 \pT}{v} ... f_\alpha, 
\eel
where $\alpha = 3S, 2S, 2D$, we rewrite  \rfn{LMdq1} and \rfn{LMdg1} in the compact form
\begin{eqnarray}
\langle w^2 \rangle_{3S} + \langle v^2 \rangle_{3S} \f{d\ln T}{d\ln \tau} +\langle v \rangle_{2D} \tau^2 T \f{d}{d\tau} \left(\f{\mu}{T} \right) &=& 0, \nn \\
\langle \f{w^2}{v} \rangle_{2D} + \langle v \rangle_{2D}   \f{d\ln T}{d\ln \tau} +  \langle 1 \rangle_{2S} \tau^2 T \f{d}{d\tau} \left(\f{\mu}{T} \right) &=& 0.
\end{eqnarray}
To proceed further it is convenient to introduce the notation
\begin{eqnarray}
A = \langle w^2 \rangle_{3S}, \,\,B = \langle v^2 \rangle_{3S}, \,\,C = \langle v \rangle_{2D}, 
\,\,D = \langle \f{w^2}{v} \rangle_{2D}, \,\,E = \langle 1 \rangle_{2S}.
\label{ABC}
\end{eqnarray}
Then, we find the proper-time derivatives of $T$ and $\mu/T$ expressed by the coefficients \rfn{ABC}
\bel{xy}
\f{d\ln T}{d\ln \tau} = \f{A E - C D}{C^2 - B E}, \quad \tau^2 T \f{d}{d\tau} \left(\f{\mu}{T} \right) = \f{D B - A C}{C^2 - B E}.
\eel
The coefficients \rfn{ABC} can be used also to express various thermodynamic derivatives. After straightforward calculations, where $T$ and $\mu$ are treated as independent thermodynamic variables,  we find
\begin{eqnarray}
\f{\p {\cal P}^{\rm eq}}{\p T} &=& \f{\p {\cal P}^{\Qp, \rm eq}}{\p T}+\f{\p {\cal P}^{\Qm, \rm eq}}{\p T}+\f{\p {\cal P}^{\G, \rm eq}}{\p T}
= \f{A}{\tau^3 T^2} - \f{D \mu}{\tau^2 T^2}, \nn \\
\f{\p {\cal P}^{\rm eq}}{\p \mu} &=& \f{\p {\cal P}^{\Qp, \rm eq}}{\p \mu}+\f{\p {\cal P}^{\Qm, \rm eq}}{\p \mu}
=  \f{D}{\tau^2 T}, \nn \\
\f{\p {\cal E}^{\rm eq}}{\p T} &=& \f{\p {\cal E}^{\Qp, \rm eq}}{\p T}+\f{\p {\cal E}^{\Qm, \rm eq}}{\p T}+\f{\p {\cal E}^{\G, \rm eq}}{\p T}
= \f{B}{\tau^3 T^2} - \f{C \mu}{\tau^2 T^2}, \nn \\
\f{\p {\cal E}^{\rm eq}}{\p \mu} &=& \f{\p {\cal E}^{\Qp, \rm eq}}{\p \mu}+\f{\p {\cal E}^{\Qm, \rm eq}}{\p \mu}
=  \f{C}{\tau^2 T}, \nn \\
\f{\p {\cal B}^{\rm eq}}{\p T} &=& = \f{C}{3 \tau^2 T^2} - \f{E \mu}{3 \tau T^2}, \nn \\
\f{\p {\cal B}^{\rm eq}}{\p \mu} &=& =   \f{E \mu}{3 \tau T}.
\label{thermder}
\end{eqnarray}
Using \rfn{thermder} we find that
\begin{eqnarray}
\kappa_1(T,\mu) &=& \left( \f{\p{\cal P}^{\rm eq} }{\p {\cal E}^{\rm eq}} \right)_{{\cal B}^{\rm eq}} 
=   \f{\p ({\cal P}^{\rm eq},{\cal B}^{\rm eq}) }{\p ({\cal E}^{\rm eq}, {\cal B}^{\rm eq} ) } 
=  - \f{A E - C D}{C^2 - B E} = - \f{d\ln T}{d\ln \tau}, \nn \\
\kappa_2(T,\mu) &=& \f{1}{3} \left( \f{\p{\cal P}^{\rm eq} }{ \p {\cal B}^{\rm eq}} \right)_{{\cal E}^{\rm eq}} 
=   \f{1}{3} \f{\p ({\cal P}^{\rm eq},{\cal E}^{\rm eq}) }{\p ({\cal B}^{\rm eq}, {\cal E}^{\rm eq} ) } 
=  - \f{D B - A C}{\tau (C^2 - B E)} = -  \tau T \f{d}{d\tau} \left(\f{\mu}{T} \right) .
\end{eqnarray} 

\subsection{Shear viscosity}
\label{s:Shear}

The shear viscosity can be obtained from the formula  $\eta = \tau ({\cal P}_T - {\cal P}_L)_{\rm NS}/2$, 
which in close-to-equilibrium situations leads to the expression
\begin{eqnarray}
\eta &=& \f{\tau}{2} \left( {\cal P}_T - {\cal P}_L \right)_{\rm NS}  \nn \\
&=&  \f{\tau}{2}   \int \f{dw d^2 \pT}{v}   \left[  \left( \f{\pT^2}{2} - \f{w^2}{\tau^2}   \right) 
\left[ k_\Q (f_{\Qp, \eq} + \delta f_\Qp +f_{\Qm, \eq} + \delta f_\Qm) + k_\G (f_{\G, \eq} + \delta f_\G) \right] \right] \nn \\
&=& 
 \f{\tau}{2}   \int \f{dw d^2 \pT}{v}   \left[  \left( \f{\pT^2}{2} - \f{w^2}{\tau^2}   \right) 
 \left[ k_\Q (\delta f_\Qp + \delta f_{\Qm} ) + k_\G  \delta f_\G \right] \right].
\label{shear1}
\end{eqnarray}
Here we used the property that the equilibrium distributions are isotropic and do not contribute to the integral \rfn{shear1}.
Using Eqs.~\rfn{df1q} and \rfn{df1g} we find
\begin{eqnarray}
\eta &=& 
- \f{\teq}{2}   \int \f{dw d^2 \pT}{v}   \left[  \left( \f{\pT^2}{2} - \f{w^2}{\tau}   \right) 
\f{w^2}{v \tau T}\right] f_{3S} ,
\label{shear2}
\end{eqnarray}
where the terms containing derivatives of $T$ and $\mu$ dropped out again due to symmetry reasons. 
Equation \rfn{shear2} can be rewritten as
\begin{eqnarray}
\eta &=& 
- \f{\teq}{2}   \int \f{d^3p}{E_p}   \left[  \left( p_x^2 - p_z^2   \right) 
 \f{p_z^2}{E_p T}\right] f_{3S} \label{shear3} \\
&=&  -\f{\teq}{2 T}   \int \f{2\pi dp\, p^6 }{E_p^2}  \int_0^\pi \sin\theta d\theta \left[  \left( \f{\sin^2\theta}{2} - \cos^2\theta   \right) 
 \cos^2\theta  \right] f_{3S}. 
\end{eqnarray}
The integral over the angle $\theta$ gives $-4/15$, hence the final result is
\begin{eqnarray}
\eta &=&  \f{4 \pi \teq}{15 T}   \int \f{dp\,p^6 }{E_p^2}   f_{3S}, 
\label{shear4}
\end{eqnarray}
which leads to Eqs.~\rfn{eta}, \rfn{etaQ} and \rfn{etaG}.

For massless quarks, \rf{etaQ} gives $\eta_\Q = 7 g_\Q \pi^2 T^4 \teq/ 450$ and $\eta_\Q = 8 g_\Q T^4 \teq/(5 \pi^2)$ for Fermi-Dirac and Boltzmann statistics, respectively. The corresponding values of pressure are: $P^{\Q, \rm eq} = 7 g_\Q \pi^2 T^4/360$ and $P^{\Q, \rm eq} = 2 g_\Q T^4 /\pi^2$, hence, for the two statistics we find $\eta_\Q = 4 P^{\Q, \rm eq} /5$. In the similar way, from \rfn{etaG} we find for massless gluons: $\eta_\G = 2 g_\G \pi^2 T^4 \teq/225$ and $\eta_\G = 4 g_\G T^4 \teq/ (5 \pi^2)$ for Bose-Einstein and Boltzmann statistics. The corresponding pressures are: $P^{\G, \rm eq} = g_\G \pi^2 T^4/90$ and $P^{\G, \rm eq} = g_\G T^4/\pi^2$, which gives again $\eta_\G = 4 P^{\G, \rm eq} /5$.

\subsection{Bulk viscosity}
\label{s:Bulk}

The bulk pressure is the difference between the average exact pressure, $({\cal P}_L + 2 {\cal P}_T)/3$, 
in the system and the reference equilibrium pressure, ${\cal P}^{\rm eq}$.
Close to local equilibrium, it can be defined by the following formula
\begin{eqnarray}
\Pi_{\rm NS} &=& \f{1}{3} \left( {\cal P}_L + 2 {\cal P}_T - 3 {\cal P}^{\rm eq} \right)_{\rm NS}   \nn \\
&=&  \f{1}{3}   \int \f{dw d^2 \pT}{v}   \left[  \left( 
\f{w^2}{\tau^2} + \pT^2 \right) 
\left[ k_\Q (f_{\Qp, \eq} + \delta f_\Qp +f_{\Qm, \eq} + \delta f_\Qm) + k_\G (f_{\G, \eq} + \delta f_\G) \right] \right.   \nn \\
&& 
\left.  \hspace{3cm}  - 3  \, \f{w^2}{\tau^2} \,  \left[ k_\Q (f_{\Qp, \eq} + f_{\Qm, \eq} + \delta f_\Qm) + k_\G  f_{\G, \eq}  \right] 
\right] \nn \\
&=& 
 \f{1}{3}   \int \f{dw d^2 \pT}{v}   \left[  \left( 
\f{w^2}{\tau^2} + \pT^2 \right)  \left[ k_\Q (\delta f_\Qp + \delta f_{\Qm} ) + k_\G  \delta f_\G \right] \right].
\label{bulk1}
\end{eqnarray}
To get the last line in \rfn{bulk1}, we have used the fact that equilibrium distributions are isotropic. It is interesting to notice that \rfn{bulk1}
can be also written as
\begin{eqnarray}
\Pi_{\rm NS}  &=& 
 \f{1}{3}   \int \f{dw d^2 \pT}{v}   \left[  \left( 
\f{w^2}{\tau^2} + \pT^2  + m^2 - m^2 \right)  \left[ k_\Q (\delta f_\Qp + \delta f_{\Qm} ) + k_\G  \delta f_\G \right] \right] \nn \\
&=& 
- \f{m^2}{3}   \int \f{dw d^2 \pT}{v}  \left[ k_\Q (\delta f_\Qp + \delta f_{\Qm} ) \right] ,
\label{bulk12}
\end{eqnarray}
where we used the Landau matching condition \rfn{LMdq} and the fact that gluons are massless.

Using the notation introduced above we find
\begin{eqnarray}
\Pi_{\rm NS}  &=& - \f{\teq}{3}   \int \f{dw d^2 \pT}{v} \left( \f{w^2}{\tau^2} + \pT^2 \right)  \left[ f_{3S} \left(\f{w^2}{v \tau^2 T} + \f{v}{\tau^2 T} \f{d\ln T}{d\ln \tau} \right) \right]
\nn \\
& & - \f{\teq}{3}   \int \f{dw d^2 \pT}{v} \left( \f{w^2}{\tau^2} + \pT^2 \right)  f_{2D}  \f{d}{d\tau} \left(\f{\mu}{T} \right) \nn \\
&=& - \f{\teq}{3}   \int \f{dw d^2 \pT}{v} \left( \f{w^2}{\tau^2} + \pT^2 \right)  \left[ f_{3S} \left(\f{w^2}{v \tau^2 T} - \f{v}{\tau^2 T} \, \kappa_1 \right) \right]
\nn \\
& & + \f{\teq}{3 \tau T}   \int \f{dw d^2 \pT}{v} \left( \f{w^2}{\tau^2} + \pT^2 \right)  f_{2D} \, \kappa_2. 
\label{bulk2}
\end{eqnarray}
Due to boost-invarince, the integral above can be done in the plane $z=0$, where $w = \pL t$, $v = E_p t$. Since $f_{3S}$ and $f_{2D}$ are isotropic, we obtain
\begin{eqnarray}
\Pi_{\rm NS}  &=& - \f{\teq}{3 \tau T}   \int d^3p \, p^2  \left[ f_{3S} \left(\f{p^2}{3 E^2_p} - \, \kappa_1 \right) \right]
 + \f{\teq}{3 \tau T}   \int \f{d^3p}{E_p} p^2  f_{2D} \, \kappa_2. 
\label{bulk3}
\end{eqnarray}
For the Bjorken flow we have $\p_\mu U^\mu = 1/\tau$, thus the Navier--Stokes relation $\Pi_{\rm NS}  = -\zeta \p_\mu U^\mu$ allows us to identify the bulk pressure as
\begin{eqnarray}
\zeta &=&  \f{\teq}{3 T}   \int d^3p \, p^2  \left[ f_{3S} \left(\f{p^2}{3 E^2_p} - \, \kappa_1 \right) \right]
 - \f{\teq}{3 T}   \int \f{d^3p}{E_p} p^2  f_{2D} \, \kappa_2. 
\label{bulk4}
\end{eqnarray}
Similarly, starting from \rfn{bulk12} we find
\begin{eqnarray}
\zeta &=&  \f{\teq m^2}{3 T}   \int d^3p   \left[ f_{2S}  \left(\kappa_1 - \f{p^2}{3 E^2_p} \right) \right]
 + \f{\teq m^2}{3 T}   \int \f{d^3p}{E_p}  f_{2D} \, \kappa_2,
\label{bulk5}
\end{eqnarray}
which leads to \rfn{zeta}.

\section{Tables of initial and final parameters}
\label{s:tables}

\begin{table}[h!]
\begin{center}
\begin{tabular}{|c|c|c|c|c|c|c|c|}
\hline
  &   $\mathcal{B}_0\left[\frac{1}{\text{fm}^3}\right]$ & $\xi_Q^0$ & $\xi_{G}^0$ & $T_0 [\text{MeV}]$ & $T_f [\text{MeV}]$ & $\mu _0 [\text{MeV}]$ & $\mu _f [\text{MeV}]$ \\\hline\hline
 \text{KT}   & 0.001 & 1 & 10 & 164 & 50 & 0 & 0 \\\hline
 \text{KT}  & 0.001 & -0.5 & 10 & 200 & 59 & 0 & 0 \\\hline
 \text{KT}  & 0.001 & -0.5 & -0.25 & 217 & 64 & 0 & 0 \\\hline
 \text{NS}  & 0.001 &  &  & 149 & 50 & 0 & 0 \\\hline
 \text{BJ}  & 0.001 &  &  & 185 & 50 & 0 & 0 \\\hline
 \text{KT}  & 1 & 1 & 10 & 164 & 54 & 246 & 55 \\\hline
 \text{KT}  & 1 & -0.5 & 10 & 199 & 60 & 202 & 47 \\\hline
 \text{KT}  & 1 & -0.5 & -0.25 & 217 & 65 & 180 & 42 \\\hline
 \text{NS}  & 1 &  &  & 130 & 54 & 281 & 55 \\\hline
 \text{BJ}  & 1 &  &  & 199 & 54 & 202 & 55 \\\hline
\end{tabular}
\end{center} 
\label{Tab1:m1cs}
\caption{Initial and final parameters for the case: $m=1$~MeV and classical statistics ($\tau_f=5$~fm). }
\end{table}
%
%
\begin{table}[h!]
\begin{center}
\begin{tabular}{|c|c|c|c|c|c|c|c|}
\hline
  &   $\mathcal{B}_0\left[\frac{1}{\text{fm}^3}\right]$ & $\xi_Q^0$ & $\xi_{G}^0$ & $T_0 [\text{MeV}]$ & $T_f [\text{MeV}]$ & $\mu _0 [\text{MeV}]$ & $\mu _f [\text{MeV}]$ \\\hline\hline
 \text{KT} & 0.001 & 1 & 10 & 165 & 60 & 1 & 1 \\\hline
 \text{KT} & 0.001 & -0.5 & 10 & 196 & 68 & 0 & 0 \\\hline
 \text{KT} & 0.001 & -0.5 & -0.25 & 214 & 74 & 0 & 0 \\\hline
 \text{NS} & 0.001 &  &  & 151 & 60 & 1 & 1 \\\hline
 \text{BJ} & 0.001 &  &  & 182 & 60 & 0 & 1 \\\hline
 \text{KT} & 1 & 1 & 10 & 161 & 52 & 342 & 223 \\\hline
 \text{KT} & 1 & -0.5 & 10 & 193 & 60 & 282 & 197 \\\hline
 \text{KT} & 1 & -0.5 & -0.25 & 215 & 68 & 242 & 167 \\\hline
 \text{NS} & 1 &  &  & 117 & 52 & 407 & 223 \\\hline
 \text{BJ} & 1 &  &  & 200 & 52 & 271 & 223 \\\hline
\end{tabular}
\end{center} 
\label{Tab2:m300cs}
\caption{Initial and final parameters for the case: $m=300$~MeV and classical statistics ($\tau_f~=~5$~fm). }
\end{table}
%
%
\begin{table}[h!]
\begin{center}
\begin{tabular}{|c|c|c|c|c|c|c|c|}
\hline
  &  $\mathcal{B}_0\left[\frac{1}{\text{fm}^3}\right]$ & $\xi_Q^0$ & $\xi_{G}^0$ & $T_0 [\text{MeV}]$ & $T_f [\text{MeV}]$ & $\mu _0 [\text{MeV}]$ & $\mu _f [\text{MeV}]$ \\\hline\hline
 \text{KT}  & 0.001 & 1 & 10 & 164 & 59 & 1 & 1 \\\hline
 \text{KT}  & 0.001 & -0.5 & 10 & 194 & 67 & 0 & 0 \\\hline
 \text{KT}  & 0.001 & -0.5 & -0.25 & 214 & 73 & 0 & 0 \\\hline
 \text{NS}  & 0.001 &  &  & 150 & 59 & 1 & 1 \\\hline
 \text{BJ}  & 0.001 &  &  & 181 & 59 & 1 & 1 \\\hline
 \text{KT}  & 1 & 1 & 10 & 160 & 53 & 391 & 223 \\\hline
 \text{KT}  & 1 & -0.5 & 10 & 191 & 59 & 325 & 202 \\\hline
 \text{KT}  & 1 & -0.5 & -0.25 & 214 & 68 & 274 & 171 \\\hline
 \text{NS}  & 1 &  &  & 103 & 53 & 494 & 223 \\\hline
 \text{BJ}  & 1 &  &  & 202 & 53 & 302 & 223
\\ \hline
\end{tabular}
\end{center} 
\label{Tab3:m300qs}
\caption{Initial and final parameters for the case: $m=300$~MeV and quantum statistics ($\tau_f=5$~fm). }
\end{table}
%
%
\begin{table}[h!]
\begin{center}
\begin{tabular}{| c|c|c|c|c|c|c|c|}
\hline
  &  $\mathcal{B}_0\left[\frac{1}{\text{fm}^3}\right]$ & $\xi_Q^0$ & $\xi_{G}^0$ & $T_0 [\text{MeV}]$ & $T_f [\text{MeV}]$ & $\mu _0 [\text{MeV}]$ & $\mu _f [\text{MeV}]$ \\\hline\hline
 \text{KT}  & 1 & 1 & 10 & 160 & 41 & 391 & 229 \\\hline
 \text{NS}  & 1 &  &  & 103 & 41 & 494 & 229 \\\hline
 \text{BJ}  & 1 &  &  & 202 & 41 & 300 & 229\\ \hline
\end{tabular}
\end{center} 
\label{Tab4:m300qs}
\caption{Initial and final parameters for the case: $m=300$~MeV and quantum statistics  ($\tau_f=10$~fm).}
\end{table}

\newpage

\bibliography{hydro_review}{}
\bibliographystyle{utphys}

\newpage

\end{document}